\documentclass[fleqn,usenatbib]{mnras}

\usepackage[dvipsnames]{xcolor}

\usepackage{newtxtext,newtxmath}

\usepackage[T1]{fontenc}
\usepackage{ae,aecompl}

\usepackage{ulem}

\usepackage{graphicx}   
\usepackage{amsmath}    
\usepackage{cite}

\usepackage{multirow}
\usepackage{booktabs}
\usepackage{tabularx}

\newpage

\graphicspath{{./figures/}}

\title[Supernovae and photoionizing feedback]{Supernovae and photoionizing feedback in spiral arm molecular clouds}

\author[T. J. R. Bending, C. L. Dobbs and M. R. Bate]{
Thomas J. R. Bending,
Clare L. Dobbs,
Matthew R. Bate
\\
School of Physics and Astronomy, University of Exeter, Stocker Road, Exeter EX4 4QL, UK\\
}

\date{Accepted XXX. Received YYY; in original form ZZZ}

\pubyear{2020}

\begin{document}
\label{firstpage}
\pagerange{\pageref{firstpage}--\pageref{lastpage}}
\maketitle

\begin{abstract}
We explore the interplay between supernovae and the ionizing radiation of their progenitors in star forming regions. The relative contributions of these stellar feedback processes are not well understood, particularly on scales greater than a single star forming cloud. We focus predominantly on how they affect the interstellar medium.  We re-simulate a 500 pc$^2$ region from previous work that included photoionization and add supernovae. Over the course of 10 Myr more than 500 supernovae occur in the region. The supernovae remnants cool very quickly in the absence of earlier photoionization, but form much larger and more spherical hot bubbles when photoionization is present. Overall, the photoionization has a significantly greater effect on gas morphology and the sites of star formation. However, the two processes are comparable when looking at their effect on velocity dispersion. When combined, the two feedback processes increase the velocity dispersions by more than the sum of their parts, particularly on scales above 5 pc.

\end{abstract}




\section{Introduction}
\label{se:intro}

Stellar feedback has long been thought to be a key factor driving both the global gas dynamics and thermal distribution of the interstellar medium, and regulating the star formation rate in galaxies. In numerical simulations, the earliest work including stellar feedback assumed supernovae feedback \citep{Katz1992}, and showed that on galaxy scales, feedback resulted in the redistribution of matter and angular momentum away from the centre of the galaxy.
Over the next decade or so most work focused on the uncertainty of how to model supernovae feedback in order to produce realistic disc galaxies \citep{Navarro1993, Thornton1998,Sommer1999,Springel2000,Efstathiou2000,Thacker2000,Thacker2001,Scan2006, Stinson2006,Governato2007,Keller2014,Kimm2015,Gentry2017}.
Typically in these simulations supernovae (SNe) needed to be inserted as kinetic energy or with cooling turned off to be effective without sufficient resolution \citep{Kay2002, Dalla2012, Hopkins2018}. The role of ionizing radiation was comparatively less well explored \citep{Haehnelt1995,Gerritsen1997,Kravtsov2000}, though \citet{Stinson2013} proposed a feedback model whereby early energy injection resembles UV radiation. This feedback was found to be more effective than SNe alone, and following on from this, numerous papers have argued that pre-supernovae feedback is essential to model galaxies (e.g. \citealt{Agertz2013,Hu2016,Hopkins2018a,Emerick2018}). 

On smaller scales, in simulations of isolated galaxies the inclusion of feedback is found to be necessary to obtain realistic Giant Molecular Cloud (GMC) and Interstellar Medium (ISM) properties \citep{Dobbs2011, Grisdale2018}. Clustering of supernovae also appears to be important to produce superbubbles \citep{Kim2017,Kim2018}. In \citet{Dobbs2011} the feedback is nominally supernovae feedback, but the main role of the feedback is that there needs to be some addition of energy to the molecular cloud which is capable of disrupting the gas. Recent work has also focused on pre-supernova feedback mechanisms of thermal and momentum feedback from HII regions \citep{Vandenbroucke2019,Jeffreson2020,Jeffreson2021}, FUV feedback \citep{Benincasa2020,Kim2017a} in addition to SNe, and the resulting GMC properties in galaxy scale simulations. 

Observations have highlighted the role of feedback prior to supernovae in dispersing the gas surrounding stellar clusters \citep{Calzetti1997,Blum2001,Hollyhead2015,  Grasha2019,Chevance2020,McLeod2021,Chevance2022}.  These show that the timescale for clusters to emerge from the embedded phase appears to be of a few to several Myr. Similarly simulations of clusters subject to ionizing feedback have shown that ionization can deplete most if not all of the gas from the natal molecular cloud (e.g. \citealt{Dale2012a,Dale2013, Colin2013,Geen2016, Ali2018, Grudic2018, Fukushima2020a}).  The implications for supernovae are then that supernovae will occur in low density, already ionized regions \citep{Walch2015,Peters2017,Kannan2020}, rather than in denser gas \citep{deAvillez2000,Dobbs2011,Lu2020}, or randomly in the ISM \citep{Joung2006,Hill2012}.

Several papers have looked at the impact of supernova feedback in HII regions, such that the supernova feedback is inserted into gas which is already ionized \citep{Walch2015,Peters2017,Butler2017,Colling2018,Haid2019,Lucas2020,Kannan2020}. Most of these simulations adopt a vertically stratified, or shearing box. \citet{Rathjen2021} model a vertical slice of the galaxy and test supernova only feedback versus including photoionization and winds initially as well, and show that the latter produces cluster and cloud properties in better agreement with observations. Generally, with the exception of \citet{Butler2017}, ionizing feedback is found to have a stronger effect on the star formation rate than supernovae, whilst \citet{Kannan2020} find that photoionziation combined with supernovae is stronger at driving outflows. Some of these studies find that the impact of a supernova event is limited due to the prior ionizing feedback, and the supernova simply acts to heat already low density gas to $\sim10^8$ K \citep{Peters2017,Lucas2020}. By contrast, when ionization is omitted, the supernova has a greater effect when the energy is instead inserted into cold gas \citep{Lucas2020}. Some of these studies however only simulate one supernovae \citep{Walch2015,Lucas2020}, or contain relatively few feedback producing sinks or stars \citep{Peters2017}.
Along a spiral arm, multiple dense gas clouds occur close together with multiple ionisation regions and supernovae occurring close together or overlapping rather than in isolation. Supernovae occurring in one cluster or cloud could potentially impact a neighbouring cloud, especially along dense spiral arms. We saw some evidence of the latter but for ionization in our previous work \citep{Bending2020}. 


Stellar feedback has also long been associated with driving turbulence in the interstellar medium (see e.g. reviews by \citealt{MacLow2004,Elmegreen2004}). Simulations by \citet{Colling2018} show that stellar feedback and shear lead to a Larson type relation \citep{Larson1981} but the velocity dispersions are a factor 2 or so too low. \citet{Seifried2018} show that SNe may have a limited effect on the velocity dispersion of the ISM if they occur outside clouds, and also that they appear to have limited impact at larger ($>$50 pc) scale lengths.

In this paper we study photoionizing and supernova feedback in clusters forming along a section of spiral arm. We investigate how effective supernova feedback is compared with ionization on these scales where multiple supernovae are occurring. We also investigate the environments where supernovae occur, and the impact of ionization and supernovae on the gas dynamics. We cover the details of the numerical methods and simulations in Section \ref{sec:methods}. Section \ref{sec:res_evo} provides a qualitative overview of our results. We then look more quantitatively at the distribution of gas in different phases, the velocity dispersion (turbulence), and cloud properties in Sections \ref{sec:gas_phases}-\ref{sec:clouds} respectively. Discussion and our conclusions are found in Section \ref{sec:conclusion}.


\section{Numerical methods}
\label{sec:methods}

The calculations presented in this paper were performed using the three-dimensional smooth particle hydrodynamics (SPH) code, {\sc sphNG}.  The code originated from W. Benz \citep{Benz1990sph-review,Benz1990}, but has since been substantially modified \citet*{Bate1995,Price2007}. The code has been parallelized using both {\sc OpenMP} and the message passing interface (MPI). The set up and initial conditions for the simulations are very similar to those presented in \citet{Bending2020}, but here we add supernovae feedback. The details of the calculations, including the initial conditions and the photoionizing feedback, are described in detail in \citet{Bending2020}, but we also provide a brief summary here.

\subsection{Details of simulations}
\label{sec:general}

The simulations in this paper model gas in a section of spiral arm. The gas is subject to a galaxy potential which comprises of logarithmic potential \citep{binney2008galactic} and a spiral component with a 2 armed spiral pattern rotating with a fixed pattern speed \citep{Cox2002}. The logarithmic potential provides a Milky Way-like rotation curve.

The gas in the simulations is subject to heating and cooling, according to \citet{Glover2007}, and self gravity. We assume a constant cosmic ray ionization rate of $\zeta_{\mathrm cr} = 6 \times 10^{-18}$ s\textsuperscript{-1}, solar metallicity and background UV heating from \citet{Glover2007}, in turn from \citet{Bakes1994} which is set to 1 $G_0$, i.e. the \citet{Habing1968} field.

Sink particles are inserted when the gas number density exceeds $1.2\times 10^4$ cm$^{-3}$ and the conditions for sink formation from \citet{Bate1995} are met, but we also allow sink formation regardless of whether these conditions are met if the number density exceeds $1.2\times 10^6$ cm$^{-3}$. The accretion radius is set to 0.78 pc. All the sink formation conditions and properties are set the same as \citet{Bending2020}. The boundaries are open, which neglects gas inflow which may become more relevant at late times. We will look into the impact of this choice in future work.

We include photoionizing feedback as described in \citet{Bending2020}. The method uses a line of sight approach, balancing the Lyman continuum flux with the recombination rate. The lines of sight are defined by every gas particle-sink particle pair and are calculated on the same timestep as the hydrodynamics simulations. Column densities along lines of sight are calculated by integration of the line integrals through each overlapping SPH particle's zone of compact support. We use a time dependent case-B recombination rate of $2.7 \times 10^{-13}$ cm\textsuperscript{3} s\textsuperscript{-1} and the on-the-spot approximation to deal with 13.6 eV photons emitted by re-combinations. Gas which is ionized is assumed to have a temperature of $10^4$ K.

We use a sampling routine to distribute stars of different masses to the sinks according to a Kroupa IMF \citep{Kroupa2001}, roughly following \citet{Geen2018}. This allows us to track massive stars, and compute each sink particle's ionizing flux. We follow photoionizing radiation of stars with masses exceeding 18 $M_\odot$. In most of our simulations we take 50\% efficiency for our ionization scheme such that half the mass in each sink particle is considered to be star mass and the other half gas, but we also run one comparison with only a 10\% efficiency (we denote this model 'Both 10\%'). We include one run with photoionizing feedback only (ionization only), which was already presented in \citet{Bending2020} and was denoted SR\_50\% in that paper. The different simulations presented in the paper are listed in Table~1. All the simulations are run for 9 to 10 Myr.

\begin{table}
\caption{List of simulations presented. The first two models were also presented in \citet{Bending2020}.}
\centering                                      
\begin{tabular}{c | c | c | c}          
\hline\hline                        
Name & ionization & Efficiency for & SNe \\ 
 & & star formation (\%) &  \\ 
\hline                                   
    No Feedback  & N &  - & Y \\ 
    Ionization only & Y &  50 & N \\      
    SNe only & N & 50 & Y \\      
    Both 50\% & Y & 50 & Y \\     
    Both 10\% & Y & 10 & Y \\   
\hline                                             
\end{tabular}
\end{table}

\begin{figure*}
\centering{\includegraphics[width=0.8\textwidth]{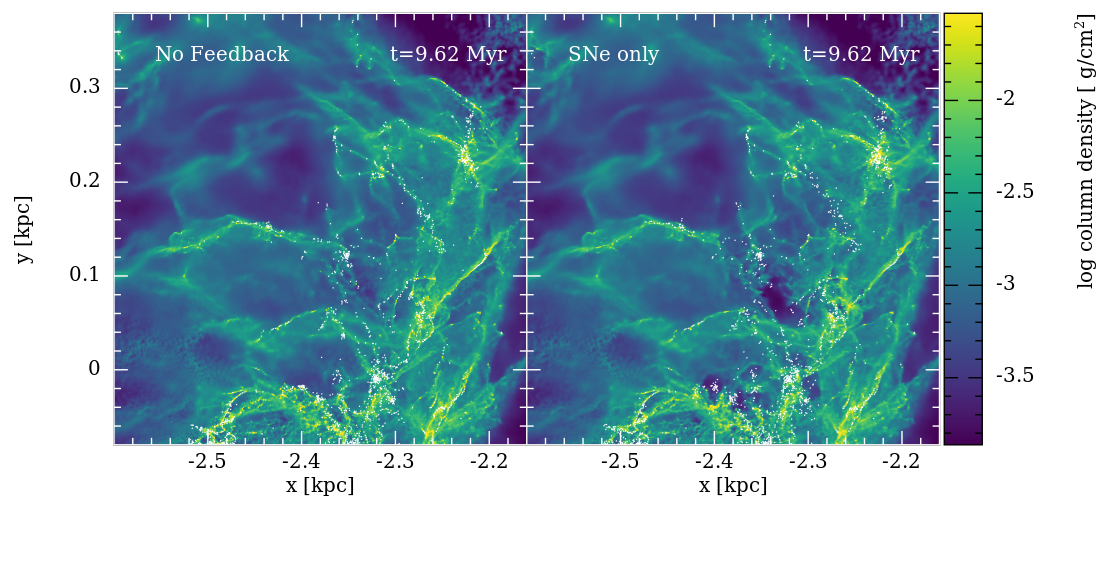}}
\centering{\includegraphics[width=0.8\textwidth]{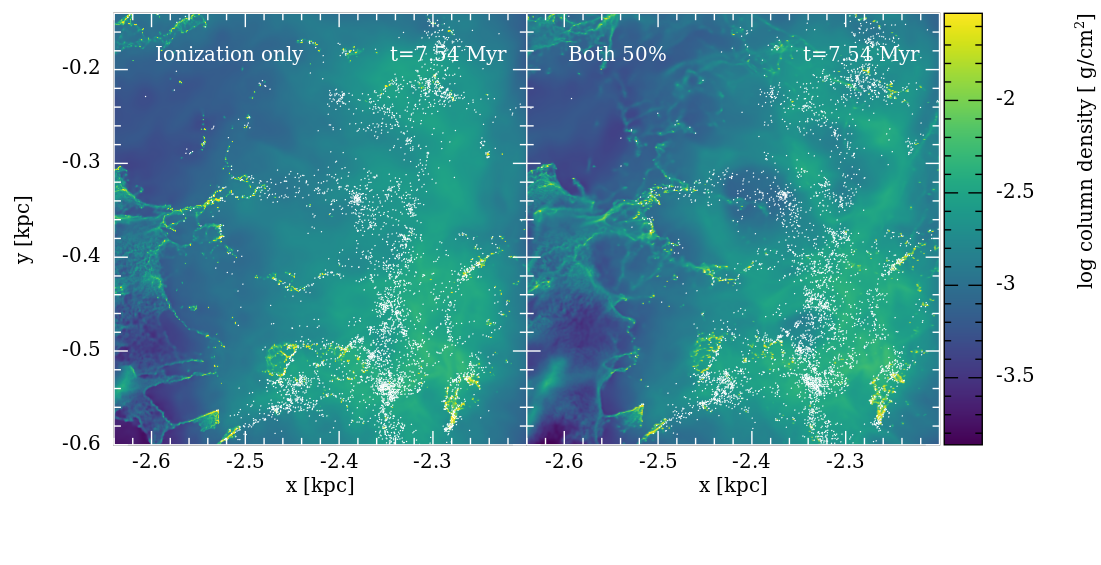}}
  \caption{The column density is shown for the models with and without supernova feedback, with no ionization (above) and with ionization (below). Sink particles are overplotted in white. Without ionization (top), the SNe have a relatively small effect, creating some low density, irregular shaped regions which are absent when SNe are not included. With ionization (lower panels), the SNe create roughly circular low density regions in warm diffuse HI. The ionization has a much greater effect than the SNe.}
  \label{fig:evolution}
\end{figure*}

\subsection{Supernovae}
\label{sec:supernovae_method}
We insert supernovae feedback largely using the same approach as \citet{Dobbs2011}, who implement supernovae directly in the pressure-driven snowplough phase, which occurs after the free expansion phase and Sedov-Taylor (adiabatic) phase which are not so well resolved. We partly use this approach so we can compare with the previous galactic scale simulations.

The main difference in the implementation is that in \citet{Dobbs2011}, particles identified as about to form a sink were used to determine the radius and density required to determine the velocity and temperature of the SNe, whereas here, since the SNe occur around sinks which have already formed, we simply take the 80 nearest gas particles.

We insert supernovae for stars of mass $>18M_\odot$, as identified from our sampling scheme. While less massive stars also undergo SN, they do so over timescales longer than these simulations. For all stars $>18M_\odot$, we determine the time after the star is formed that it will undergo a supernova event, using the SEBA program \citep{Portegies2012,Toonen2012}. Then at the start of each timestep we determine if there are any massive stars which have exceeded their lifetimes. 



To input each supernova, firstly we identify the radius ($R_{\mathrm{shock}}$) which contains the closest 80 SPH particles to the sink where a SN event is occurring. If multiple supernovae occur at similar times, in a single sink, we simply increase the energy by a factor of the number of SN ($N_{\mathrm{SN}}$), however, this occurs in only a very small number of cases. We then approximate the age of the SNR ($t$) at this shock radius following \citet{Ikeuchi1984}, as described in the appendix of \citet{Dobbs2011}. Following \citet{Ikeuchi1984} and \citet{Cioffi1988} we determine the temperature and velocity of the SN remnant, again as described in our previous work. 

\subsection{Evolution of simulations}

\begin{figure*}
\includegraphics[width=\textwidth]{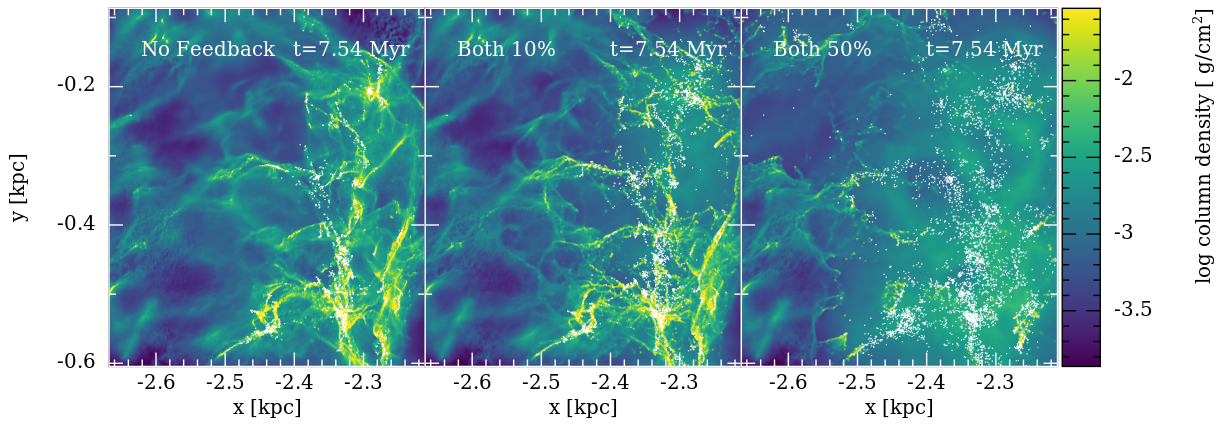}
  \caption{The column density is shown for the models with No feedback, and with both  supernova feedback, with ionization, with the 10\% efficiency (middle panel, `Both 10\%' model) and 50\% efficiency (right, `Both 50\%' model). The efficiency reflects the amount of gas which is assumed to be converted to stars within a sink, and which leads to stellar feedback. The `Both 10\%' model appears more realistic compared to the other models, as the feedback does not entirely disrupt the spiral arm, but is having some effect particularly in the upper part of the spiral arm.}
  \label{fig:efficiency}
\end{figure*}

\citet{Dobbs2011} find that this numerical approach agrees well with analytic solutions of supernovae remnants as long as the timesteps are not too long. \citet{Saitoh2009} also point out the dangers of neighbouring particles being on very different individual timesteps when dealing with explosive problems. We implement a timestep limiter to ensure timesteps are small enough for gas affected by SNe. We alter the global timestep for the simulation according to the radius of the supernova remnant divided by the sound speed of the gas. We tested this method by comparing our results using 10 times shorter timesteps than predicted by our algorithm, and checking that there was no difference in the temperature evolution. This is probably a more conservative approach compared to most methods but ensures that problems with timesteps are avoided, and we can follow supernovae where the temperatures are as high as $10^8$ K. 

As shown in Table~1, we study the effects of adding SNe to two simulations, one with ionization (labelled `Both 50\%') and one without ionization (labelled `SNe only'). We also include SNe in our 'Both 10\%' model, with the lower efficiency for forming stars. In this case there are fewer massive stars resulting in less photoionization, and fewer SNe. For most of the results, we compare the first four models, i.e. we test the impact of SNe with photoionizing feedback and without. However, as shown in \citet{Bending2020}, the photoionization has a big impact on the gas so in the model we label `Both 10\%', we examine what happens when SNe occur with a lower amount of photoionization. 


\subsection{Initial conditions}

The initial conditions for all the simulations presented in this paper are the spiral-arm region denoted SR in \citet{Bending2020}, they are are extracted from galaxy scale simulations \citep{Dobbs2013} and scaled up in resolution. The region is 500 pc\textsuperscript{2} in the plane of the galaxy and is unlimited in the z-direction. The original galaxy scale simulations included the same heating and cooling, H$_2$ and CO chemistry, and underlying spiral galaxy potential as described in Section~\ref{sec:general}. They also included a simple feedback prescription. The original particle mass is $\approx 300 M_\odot$ and after increasing the resolution the particle mass is $\approx 1 M_\odot$. The resolution is increased by splitting the particles and positioning new particles according to the SPH smoothing kernel, as described in \citet{Bending2020}. These initial conditions benefit from a realistic network of neighbouring GMCs, situated along a spiral arm, but are typically lower in resolution compared to initial conditions from spherical turbulent cloud or turbulent box models.

\section{Results}
\label{sec:results}
\subsection{Evolution of simulations}
\label{sec:res_evo}

In all our models, the spiral arm contains molecular clouds which undergo localised gravitational collapse to form sink particles.
We show snapshots from our 4 main models (No Feedback, SNe only, Ionization only, Both 50\%) in Figure~\ref{fig:evolution}. We compare the evolution with (right panels) and without supernova feedback (left panels), and with (lower panels) and without (upper panels) photoionization. The supernovae start occurring from times of around 5 Myr in the simulations. We show the models without photoionization at slightly later times for two reasons; firstly that the SNe bubbles are less obvious at earlier time, and secondly that in the ionization only model the ionization continues past when supernovae would occur, which means that at the latest times, some differences in the structure result from the continuing ionization rather than whether SNe are present. 

We find that in both cases supernovae feedback does not make that much difference to the evolution of the gas on scales $>$50-100 pc. In the case without photoionization (top panels), SNe create low density holes in the gas, most visible near to the clusters emerging from the spiral arm rather than the spiral arm clouds themselves. 
In the latter region, the cooling of the gas may be very efficient such that SNe quickly cool before having significant impact on the structure. Some are clearly associated with clusters (e.g. at $x\sim -2.4$ kpc, $y\sim0.02$ kpc), whilst the large bubble at $x\sim -2.32$ kpc, $y\sim0.14$ kpc is primarily from SNe occuring in sink particles to the top left, but the hot gas has channelled through to this region. From $\sim8$ Myr to the latest time (10.4 Myr), the holes seen in Figure~\ref{fig:evolution} top panel simply grow slightly larger. 

\begin{figure*}
  \includegraphics[width=\textwidth]{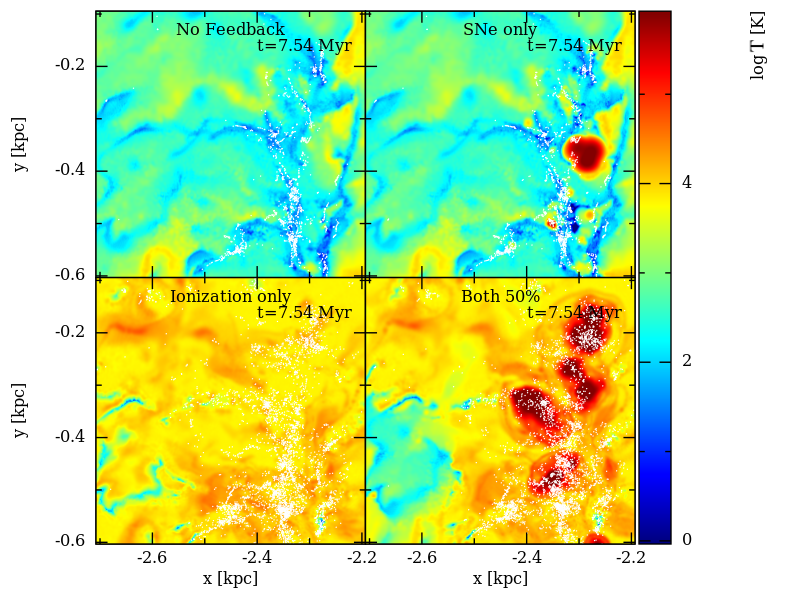}
  \caption[Temperature cross-section for runs with and without supernovae]{Temperature cross-section (slice), centred on z=0, for runs with and without feedback. Both the right hand panels have been subjected to around \textcolor{red}{80} SNe. Sink particles are overplotted in white. Ionization leads to gas at temperatures of $\sim10^4$ K, which is widespread in the models with ionization. SNe lead to gas above $10^6$ K, which is located in numerous bubbles in the model with both forms of feedback, but less prolific when only SNe are included.}
  \label{fig:temperature}
\end{figure*}

From Figure~\ref{fig:evolution} lower panels, we see that in our fiducial models (with 50\% efficiency), photoionization has a much greater effect on the gas than the SNe. The photoionization heats up the gas in the spiral arm and produces warm diffuse HI. The photoionization also obviously occurs earlier than the SNe. SNe have a noticeable impact on the gas slightly earlier compared to with no photoionization, but the morphologies of the SNe bubbles are very different. With photoionization, the SNe create low density, roughly spherical bubbles and the gas in the regions is diffuse and warm. By contrast without photoionization, the SNe produce irregular shaped bubbles with a clearly higher density contrast to the surrounding gas.  


In both cases we find that the supernovae feedback does not make that much difference to the evolution of the gas. The energy injected by SNe is resisted very effectively by dense gas, it is channelled into low density regions, including above and below the plane of the disc, and/or creates them if they do not already exist. This is the same behaviour seem by other authors in simulations with both photoionization \citep{Walch2015,Lucas2020} and stellar winds \citep{Rogers2013}.



In Figure~\ref{fig:efficiency} we show the structure of the gas in the simulations with no feedback, and with both photoionizaton and SNe feedback with different efficiencies. For the Both 10\% model, with the lower efficiency, the photoionization has a lesser effect, though the photoionization is still having a greater effect compared to supernovae. This model is likely more realistic than the 50\% model, as much more of the dense gas remains, particularly in the lower region of spiral arm. In the upper region, the morphology of the gas is still more similar to the Both 50\% model, in that supernovae are producing large low density bubbles in the diffuse gas. The difference compared to the Both 50\% model is that there is high density, cold gas still left even where feedback is occurring.     

\begin{figure}
  \centering
  \includegraphics[width=\columnwidth]{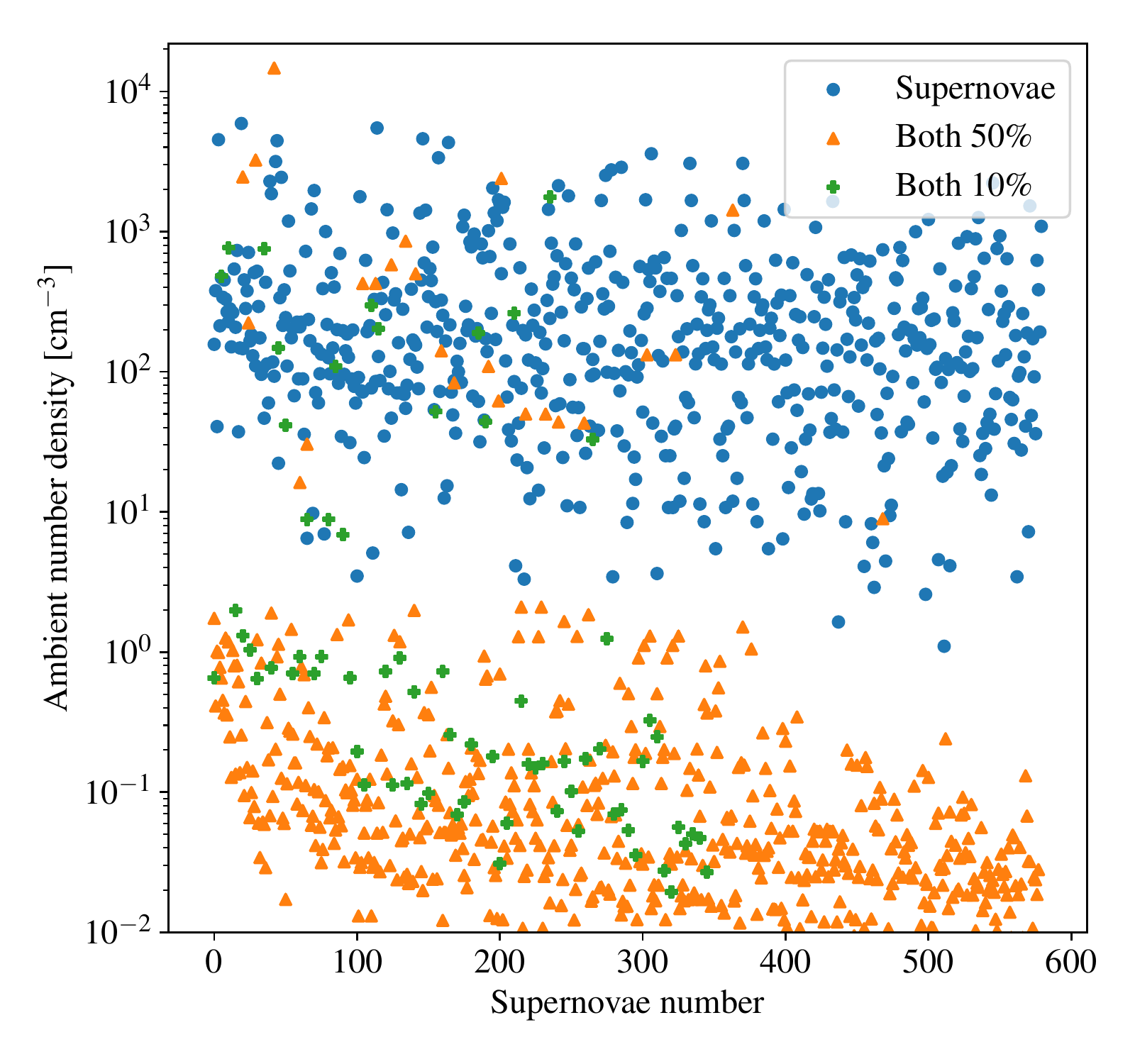}
  \vspace{-0.5cm} 
  \caption{Ambient density at sites of SNe for the first 580 SNe to occur in runs 3 and 4, arranged by the order in which they occur. The blue points are for the run without pre-SN feedback (SNe only), the orange for the run with both forms of feedback (Both 50\%), and the green for the lower feedback run (Both 10\%).}
  \label{fig:en_ratio_dens}
\end{figure}

Figure~\ref{fig:temperature} shows the temperature cross-section at the galactic mid-plane for the four simulations at the time of 7.5 Myr. Again we see a bigger difference between the simulations with and without photoionization, compared to with and without supernovae. We see that much of the gas in the models with photoionization is around $10^4$ K. We do see one clear bubble of $>10^6$ K in the supernovae only case, but generally in the supernovae only case, the SNe sites cool rapidly. In the runs with both forms of feedback, we see more hot gas and superbubbles form around the sites of SNe. Here the gas is unable to cool and we see extended regions of hot gas of size $\sim100$ pc, corresponding to the bubbles we see in the diffuse gas in the column density images (Figure~\ref{fig:evolution} lower right). For the Both 10\% model (not shown), the temperature distribution is intermediate between the No Feedback and Both 50\% models, as would be expected with less photoionizing feedback and fewer SNe. The temperature cross-sections are similar at later times, up to $\sim 10$ Myr. The supernovae only model still just shows one or two isolated bubbles of hot gas at later times. 


The effects of SNe vary a great deal depending on whether the ISM has already been affected by photoionization. In Figure~\ref{fig:temperature}, gas heated by SNe without previous feedback is able to cool quickly around all but one cluster. However, when SNe occur inside HII regions the lower density gas is unable to cool and creates large superbubbles at temperatures above 10$^6$ K. Over time these superbubbles begin to overlap and begin to form a network of $>10^6$ K gas.

One consideration for the models with photoionization is that in the model without supernovae from \citet{Bending2020}, the sink sub-grid model did not follow the massive star ages for SNe. This means that photoionizing sinks continue to emit the Lyman flux of massive stars beyond their lifetimes, and thus the amount of ionization is overestimated at later times. In the run with supernovae and photoionization however, the Lyman flux of sinks is reduced appropriately after each SN event. This doesn't appear to effect the densities much (e.g. comparing Figure~\ref{fig:evolution} lower panels) until the latest times in our simulation. However there is more of an impact on the temperature distributions, as we show in the next Section.





\begin{figure*}
  \includegraphics[width=0.46\textwidth]{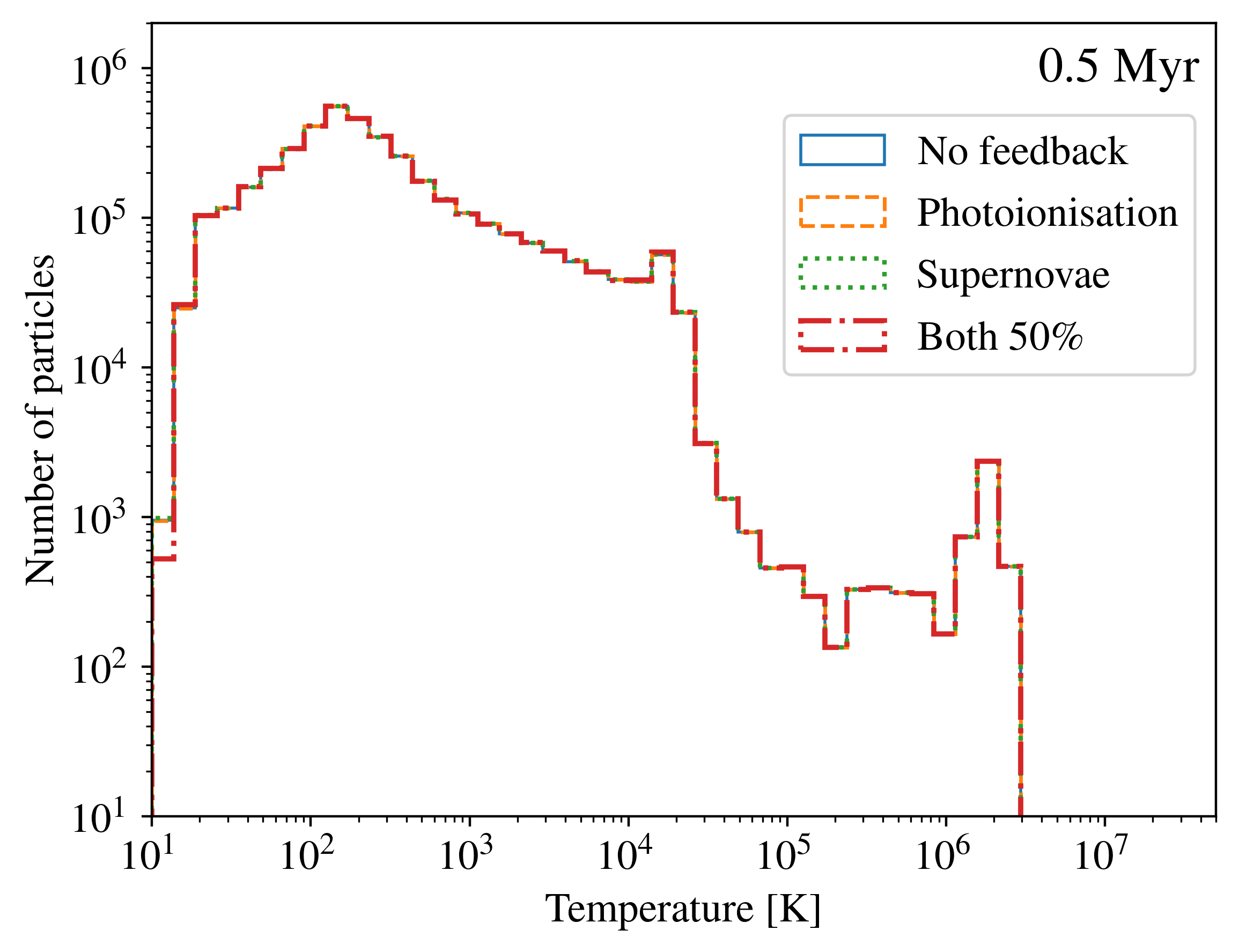}
  \includegraphics[width=0.46\textwidth]{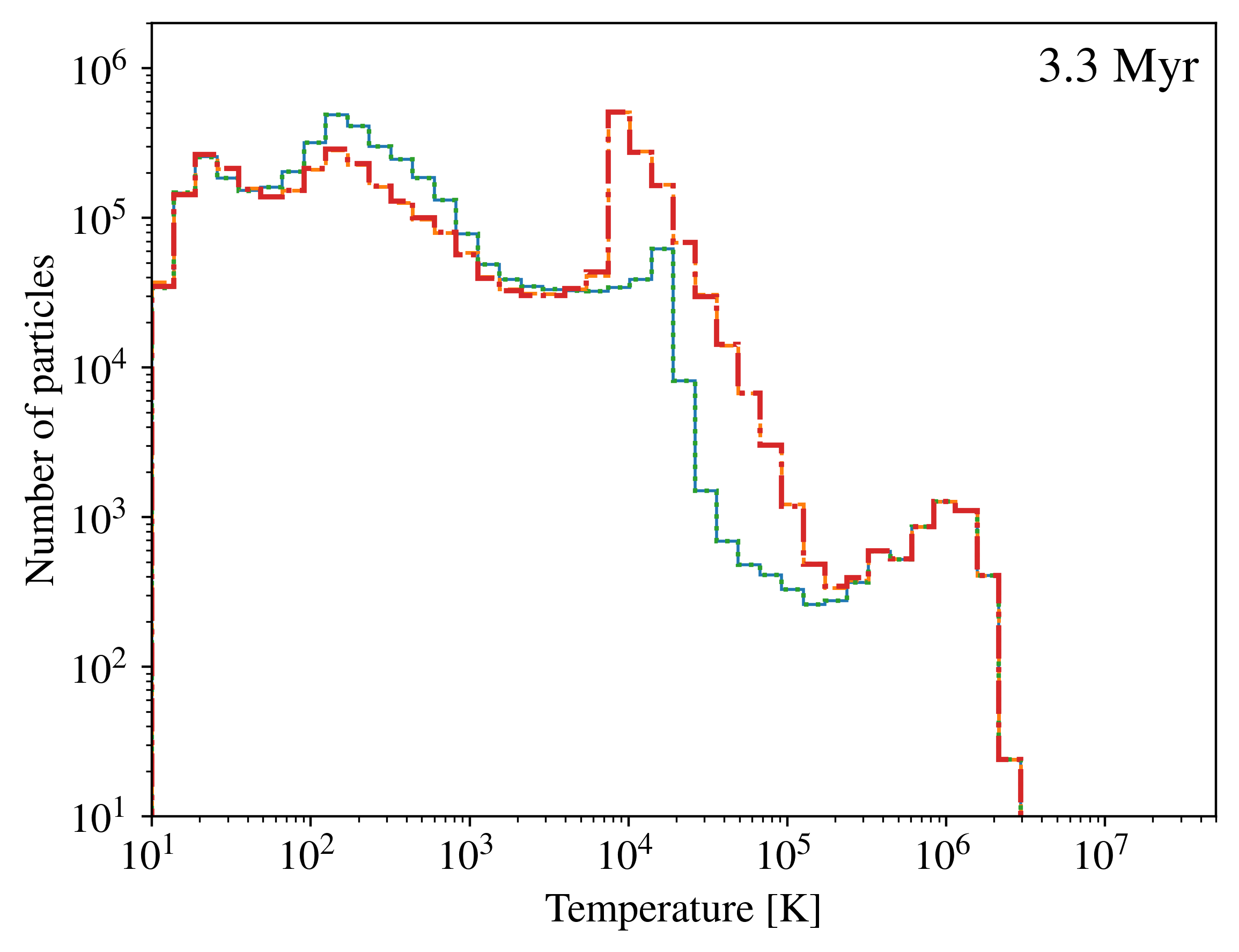}
  \includegraphics[width=0.46\textwidth]{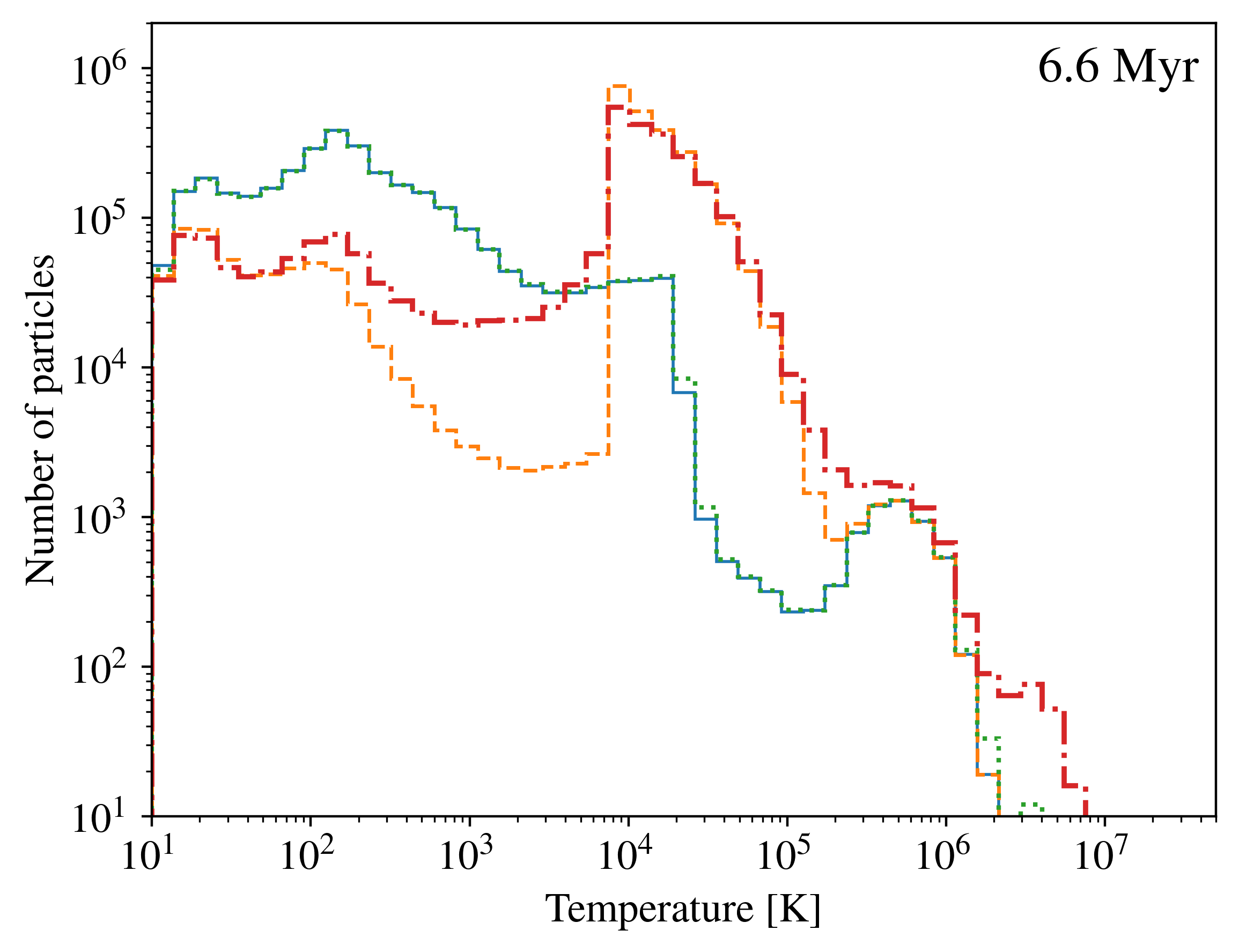}
  \includegraphics[width=0.46\textwidth]{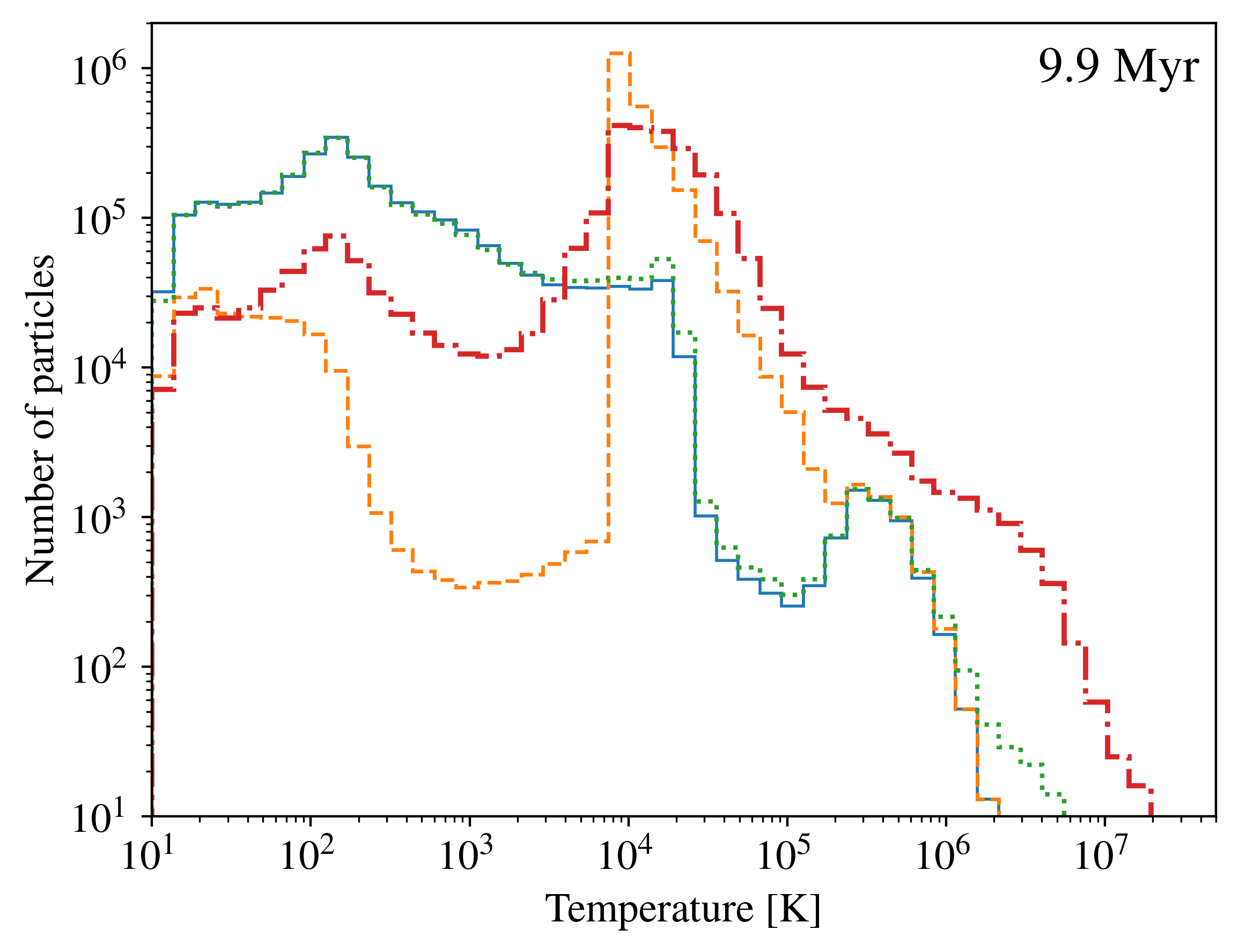}
  \caption[Temperature PDFs with and without supernovae]{Temperature PDFs for all gas particles at four timestamps in the simulations. The time of 0.5 Myr shows the inherited PDF from the galaxy scale simulations \citep{Dobbs2013}. At 3.3 Myr this is still before the first SN so this shows the initial impact of photoionization. The temperature at 6.6 Myr shows the impact of significant photoionization (heating gas to $10^4 K$) and the effect of tens of SNe (heating gas to $10^7$ K). After 9.9 Myr around 1000 SNe have occurred.}
  \label{fig:temp_PDF}
  \includegraphics[width=0.96\textwidth]{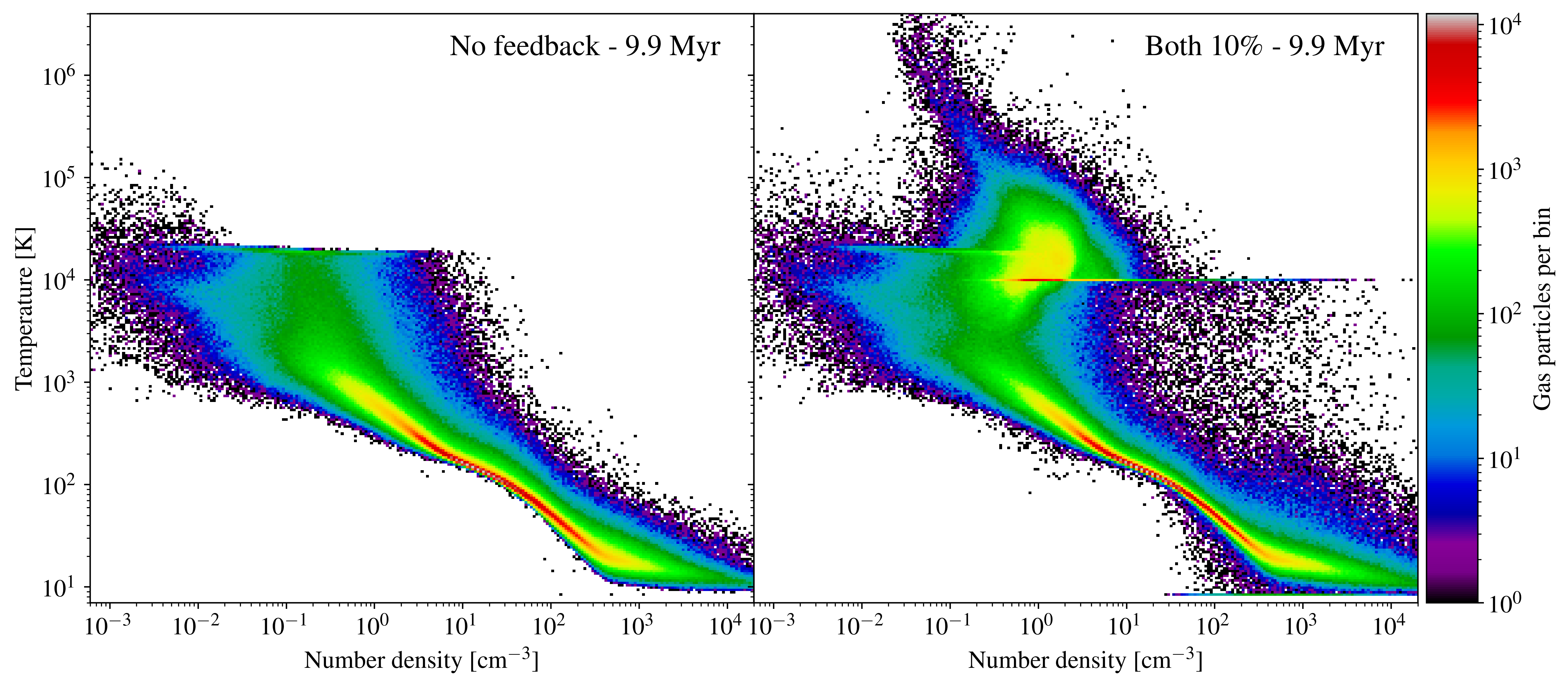}
  \caption{Density-temperature phase diagrams for the no feedback run (left) and the run with both ionizing and SNe feedback with 10\% efficiency (right). The full range of both temperature and density space have been divided up into 250 bins each. The number of SPH gas particles in each bin is plotted on the colour axis, making the bins mass weighted. In the left hand panel molecular gas can be found in the tail to the bottom right and the warm neutral medium just above $10^4$ K is seen as a nearly horizontal feature. In the right hand panel the warm ionised medium and the hot ISM can also be seen.}
  \label{fig:phase_diag}
\end{figure*}

As mentioned in the Introduction, some models have inserted supernovae primarily in high density gas, whilst others have inserted them randomly. Simulations that have consistently followed ionization and supernovae have tended to simply model one supernovae, or followed a small number of supernovae, in a single molecular cloud. 
Here we can follow many supernovae and so can determine the range of densities they occur at (with the exception of SNe from massive stars which are dynamically ejected; we don't follow these since we do not resolve cluster dynamics). We show the distribution of densities at sites of SNe in Figure \ref{fig:en_ratio_dens} for the models with just supernovae, and with both photoionization and supernovae. The densities are approximated as the average density of the 80 particles into which the SNe energy is injected. The ambient density in which SNe occur varies by 4 dex in run 2 and 7 dex in run 4. As is expected the SNe in the run with photoionization occur at much lower densities on average. This is in agreement with past work which similarly find that supernoave occur in gas orders of magnitude less dense if photoionization is included \citep{Peters2017,Kannan2020}. Supernovae do still occur in dense regions in the simulation with photoionization, but this is clearly atypical.

\subsection{Gas phases}
\label{sec:gas_phases}
To more quantitatively compare the effects of the different feedback processes on the thermal properties of the ISM, we show temperature PDFs of the gas at four different times in Figure~\ref{fig:temp_PDF}, for the four fiducial models. For most of the models, we see an indication of peaks in the PDFs occurring at around 100 and $10^4$ K. These are indicative of the cold and warm HI phases of the ISM (see also \citealt{Dobbs2008}). After 0.5 Myr, the PDFs of the different simulations are more or less identical and reflect the initial conditions taken from the global galaxy simulations. Feedback in the galaxy scale simulations included feedback which heated some gas to high temperatures. By 3.3 Myr, the models with photoionization start to show more gas at $10^4$ K, which is the temperature ionization heats the gas to.

By 9.9 Myr, the photoionization only model shows a very large peak at $10^4$ K and a dearth of gas below $10^4$ K. However the differences at lower temperatures ($10^2$ to $10^4$ K) are unlikely to be realistic and are a consequence of allowing photoionization to continue past the point supernovae would occur. This results in a higher Lyman flux, with the most massive and shortest stars in the simulation due to undergo supernovae also being those with the highest fluxes. By contrast, the model which includes both photoionization and supernovae looks more realistic. This model does not show such a drop off at intermediate temperatures, and still has a peak at 100 K. 

At the later times of 6.6 and 9.9 Myr, we also see that there is gas lying at temperatures $>10^6$ K which is not present in the models without supernovae. There is also clearly much more gas at temperatures $>10^6$ K in the model with both photoionization and supernovae. This again suggests that the cooling is more efficient in the model with supernovae only, where the supernovae occur in dense gas, compared to the model with both feedback, where supernovae occur in low density regions and gas remains hot. The temperature PDF for the `Both 10 \%' model lies between those for `No feedback' and `Both 50\% models.

Figure~\ref{fig:phase_diag} shows the concentration of SPH particles in density-temperature phase space. The plot compares the gas from the `No feedback' run and the `Both 10\%' run at the latest time shown in Figure~\ref{fig:temp_PDF}. The `No feedback' run in the left panel shows a two-phase ISM, with gas preferentially at cold and warm ($10^4-2\times10^4$ K) temperatures.
For the `Both 10\%' run in the right panel we see a multi-phase ISM, the ionisation contributing to gas at $10^4$ K from moderate to higher densities, and gas above $2\times 10^4$ K due to supernovae. The transition between the warm ionised medium and the hot ionised medium is continuous.

\subsection{Gas dynamics}
\label{sec:veldisp}
In this section we look at the effect of the different feedback mechanisms on the gas dynamics and their ability to drive motions in the gas on different scales. To do this, we calculate the velocity dispersion within spheres of different radii within our simulation. The spheres are chosen randomly, the centres between the $x$ and $y$ dimensions of a simulation at a given snapshot, and within $|z|<10$ pc. We take 1000 spheres for a given radius and calculate the average dispersion across our 1000 samples. We note that for small radii, only a small number of particles are typically selected, likely from different environments for each sphere, so the dispersions tend to be quite noisy.

\begin{figure}
  \includegraphics[width=\columnwidth]{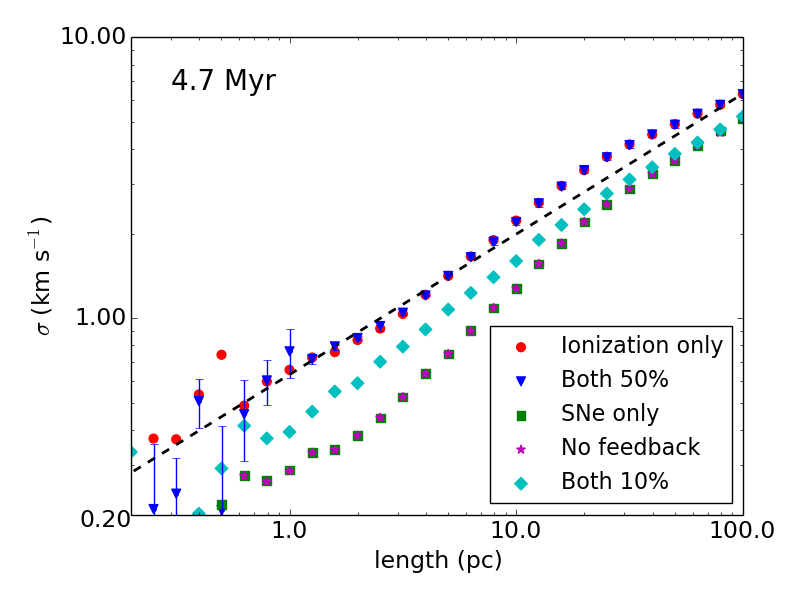}
 \includegraphics[width=\columnwidth]{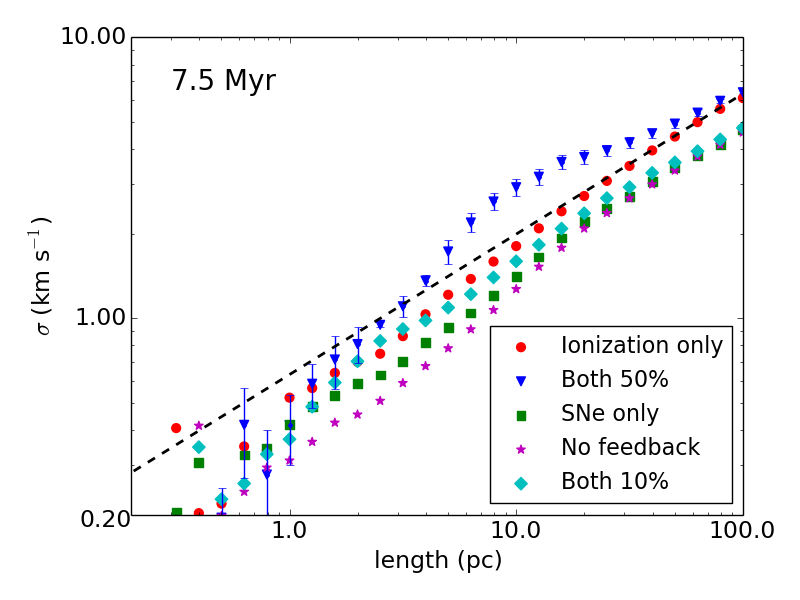}
  \includegraphics[width=\columnwidth]{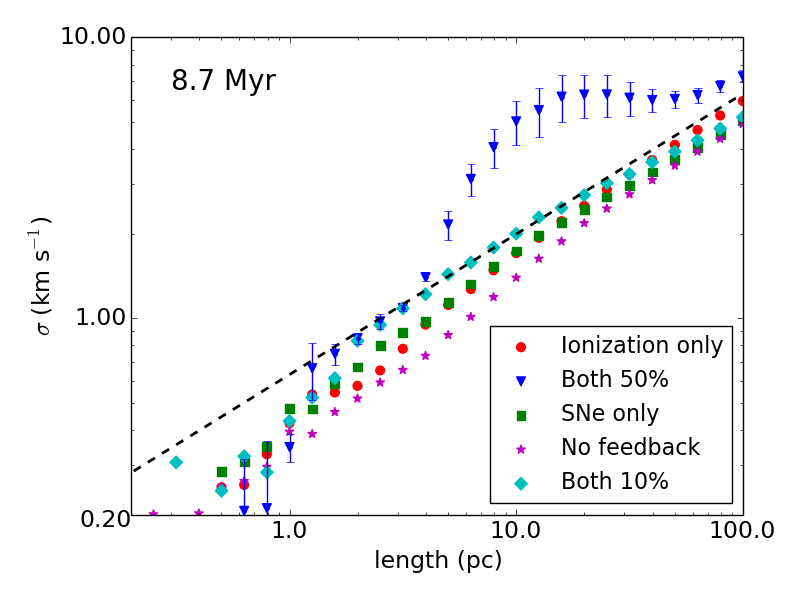}
  \caption[The velocity dispersion]{The velocity dispersion (calculated as described in the text) is shown versus different length scales at times of 4.7 Myr (top panel), 7.5 Myr (middle panel) and 8.7 Myr (lower panel) for the different models. At the earlier time, SNe are not yet effective, but photoionization has still increased the dispersion. The models with both photoionization and SNe show the highest dispersions at the last time frame, though the magnitudes are more consistent with observations in the `Both 10\%' model. The dashed line shows $\sigma\propto l^{0.5}$. Error bars show the uncertainty for the `Both 50\% model', error bars are similar or smaller (particularly at large lengths at the 8.7 Myr time) for other models.}
    \label{fig:vdisp}
\end{figure}

Figure~\ref{fig:vdisp} shows the velocity dispersion for all the different models at times of 4.7 Myr (top), 7.5 Myr (centre) and 8.7 Myr (lower). To some extent, and particularly in the No Feedback case, the velocity dispersions will be set from the initial galaxy simulation and dynamics of the gas.  At the earlier time, the gas dispersion is higher at all radii when photoionization is included. The SNe have not yet occurred in the simulation so the 'Both 50\%' model is identical to the photoionization model at most radii (and likewise the `SNe only' model compared to the `No Feedback' model). For the `Both 10\%' model the photoionization has a lower effect on the dispersions, and little effect at high length scales.

At 7.5 Myr, the model with photoionsation and SNe shows the highest velocity dispersion, with these feedback processes increasing the velocity dispersion from $\sim$ 1 km s$^{-1}$ to $\sim$ 3 km s$^{-1}$ at 10 pc length scales. There is less increase in the velocity dispersion at larger radii. At the latest time, the velocity dispersion is driven up higher at 10-50 pc suggesting the SNe continue to drive velocities up to larger scales. Reaching higher dispersions at longer ($>$ 50 pc) length scales may take longer than our simulations run for.
We also see that the `Both 50\%' model now has higher velocity dispersions compared to the other simulations The models with supernovae, and photoionization on their own show slightly higher dispersions than the no feedback model. Overall though we find that both photoionization and supernovae contribute to driving motions, or turbulence, and photoionization seems to have slightly greater effect than supernovae. 

We also show on Figure~\ref{fig:vdisp} a Larson type relation \citep{Larson1981} of $\sigma =10^{-0.2} l^{0.5}$ as observed for Milky Way clouds and complexes by \citet{Luong2016} (see also e.g. \citealt{Solomon1987,Brunt2003,Heyer2004}). We note that unlike the observations here we have not necessarily used GMCs, but consider all the gas. 
At later times, the velocity dispersions exceed the typical observed values in the `Both 50\%' model, likely indicative of feedback being too strong in this model. 
The velocity dispersion in the more realistic `Both 10\%' model matches the observed relation quite well at most length scales, whilst the SNe only and No feedback models tend to have slightly lower velocity dispersions.
We note though that in all cases the feedback is far from equilibrium, the photoionization and SNe do not occur from the start of the simulation, and SNe occur only at later times, so we can only see the relative effects of photoionization versus SNe, and that the models with only photoionization or SNe have difficulty reaching or maintaining high velocity dispersions. We plan to show models where feedback is relatively uniform throughout the duration of the simulations in future work. 



\begin{figure*}
 \includegraphics[width=0.45\textwidth]{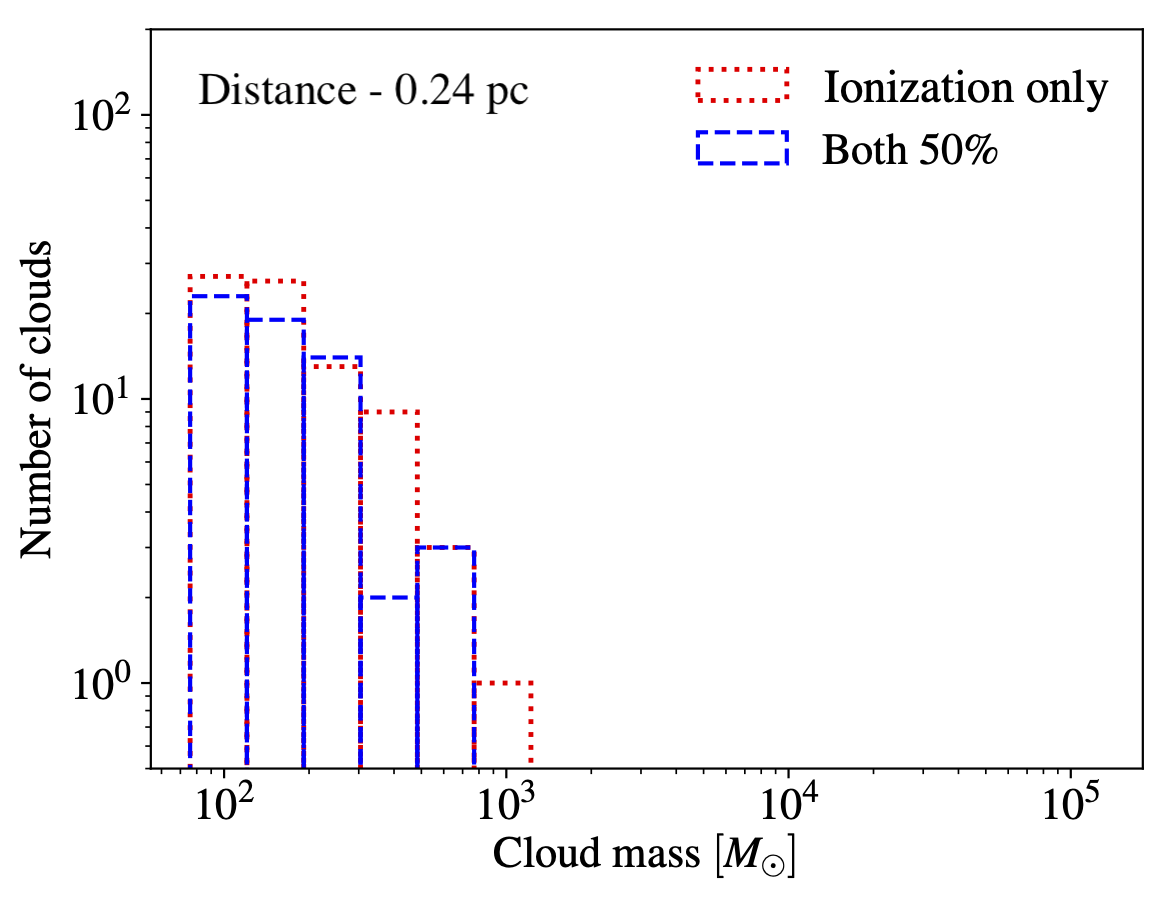}
  \includegraphics[width=0.45\textwidth]{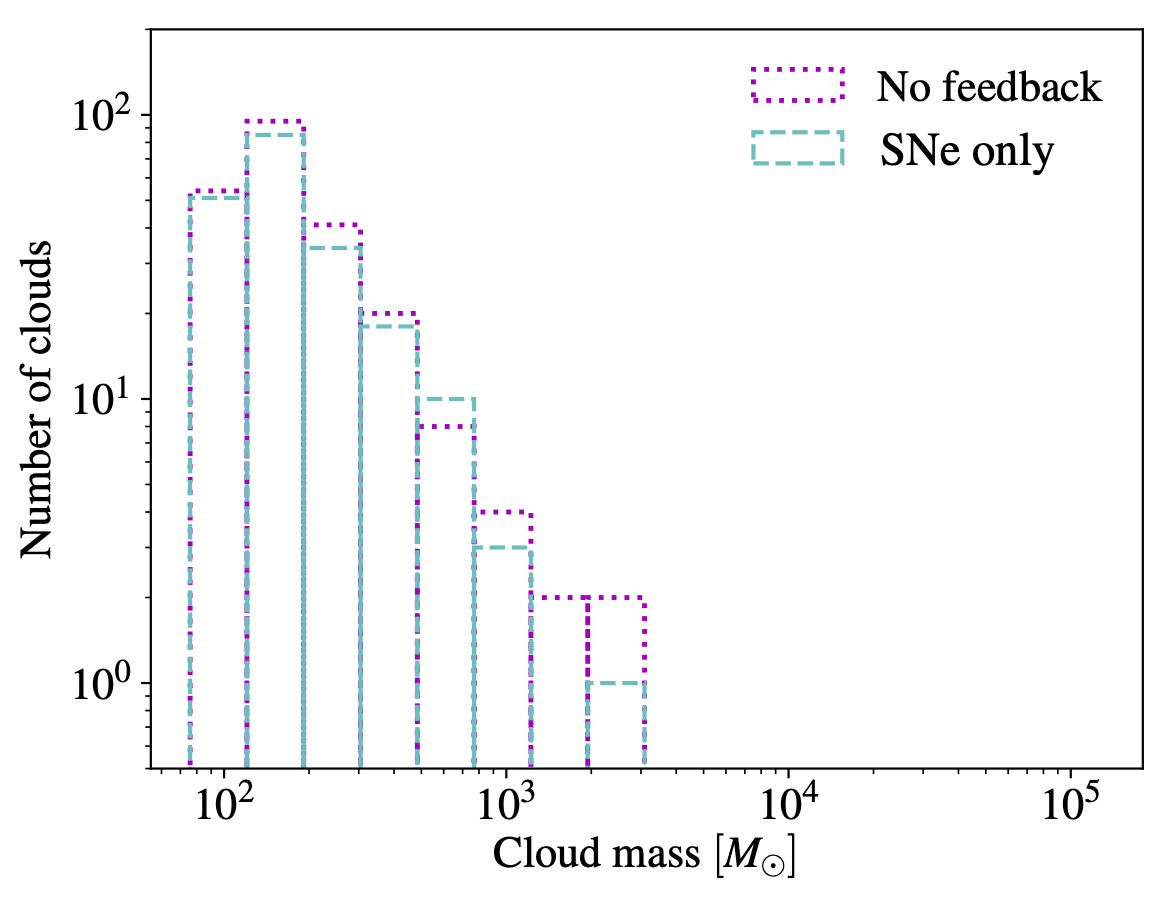}
 \includegraphics[width=0.45\textwidth]{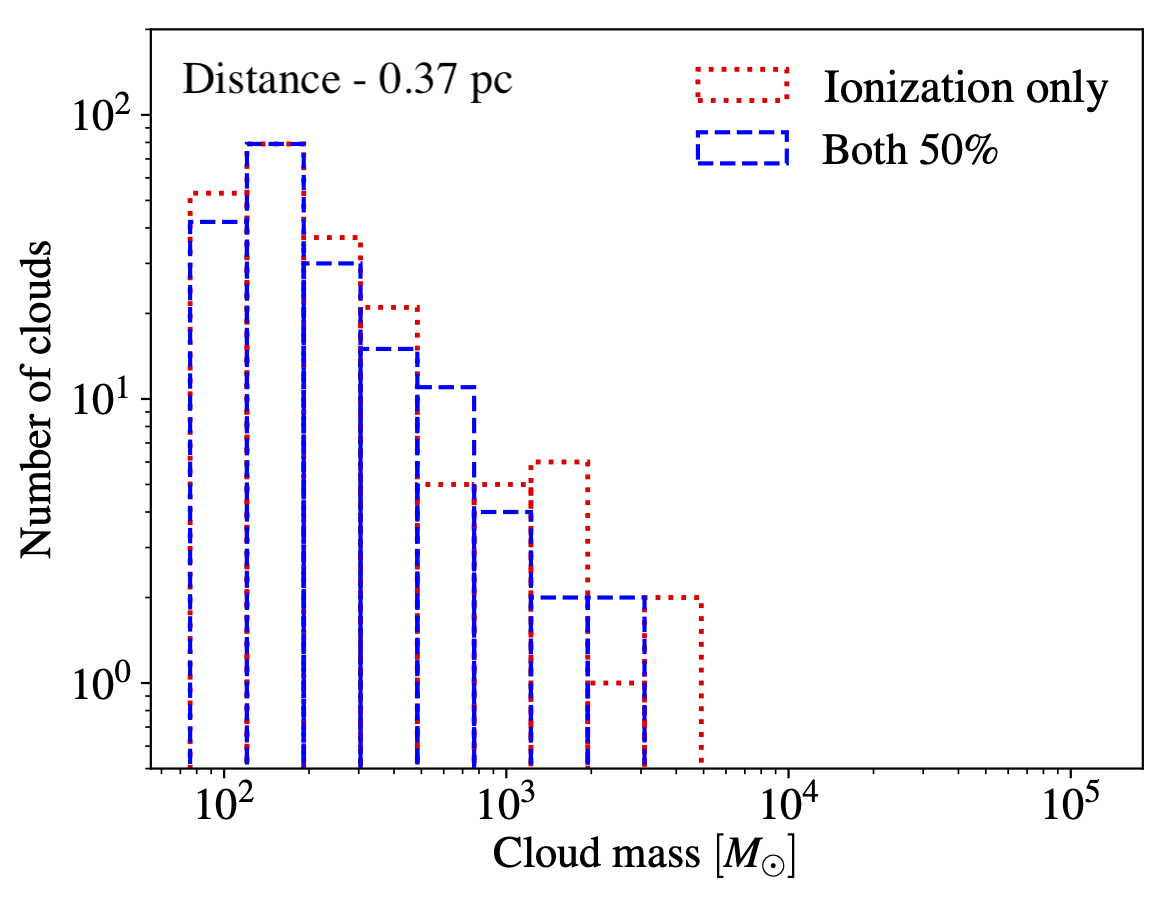}
    \includegraphics[width=0.45\textwidth]{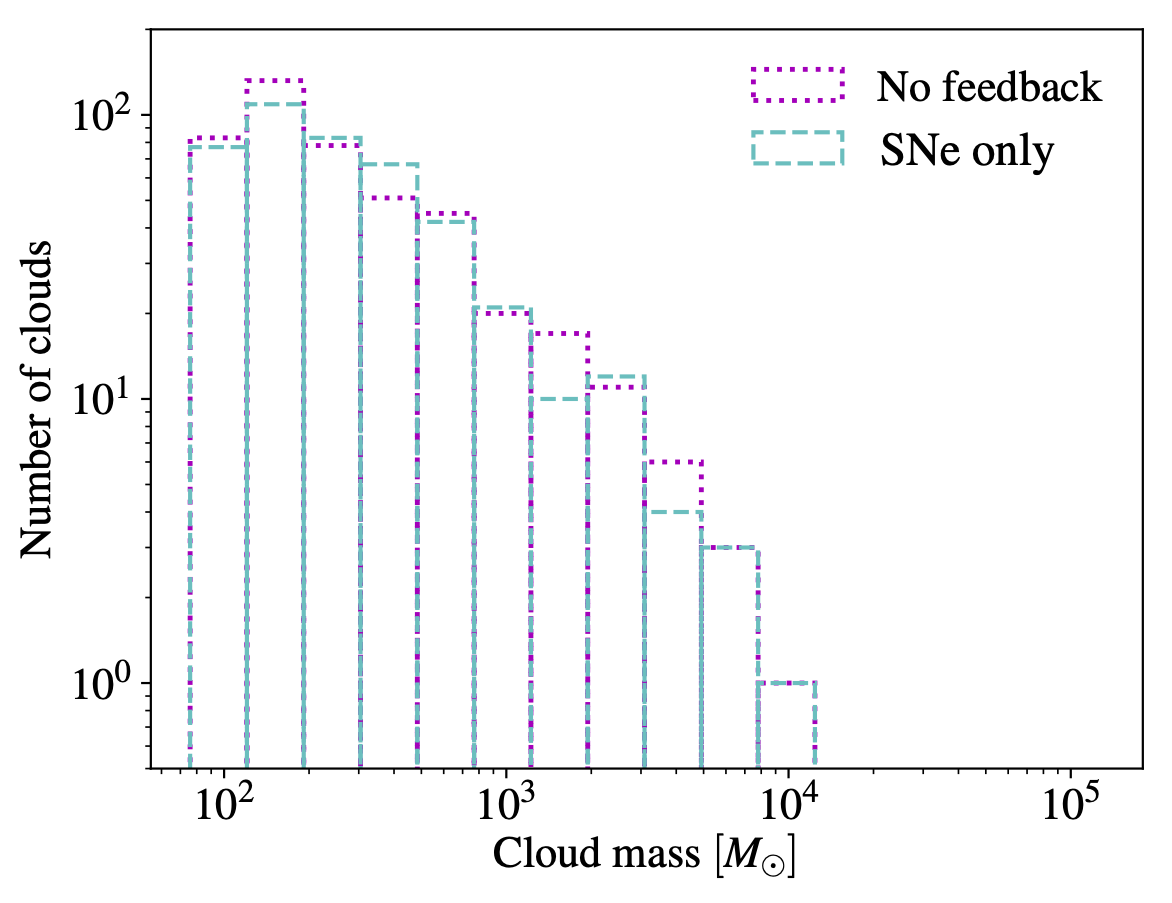}
\includegraphics[width=0.45\textwidth]{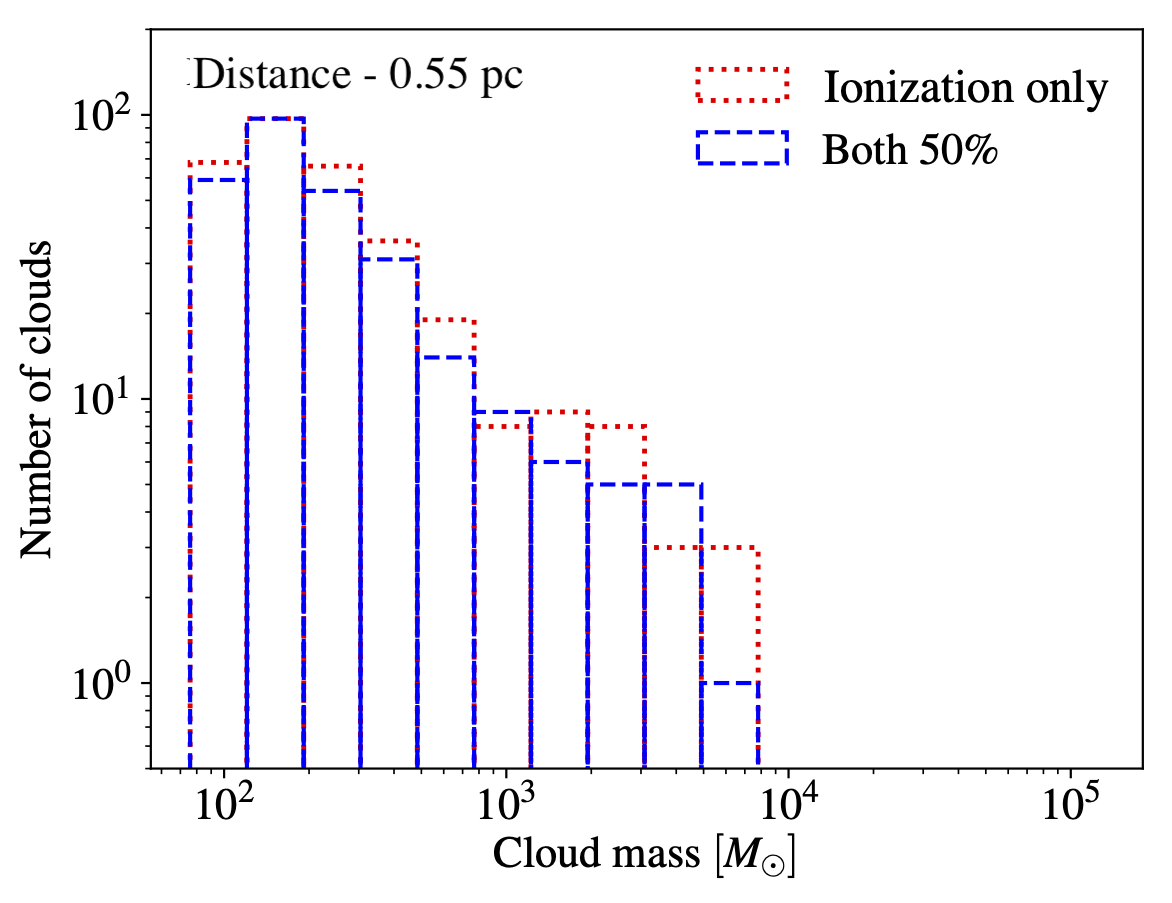}
    \includegraphics[width=0.45\textwidth]{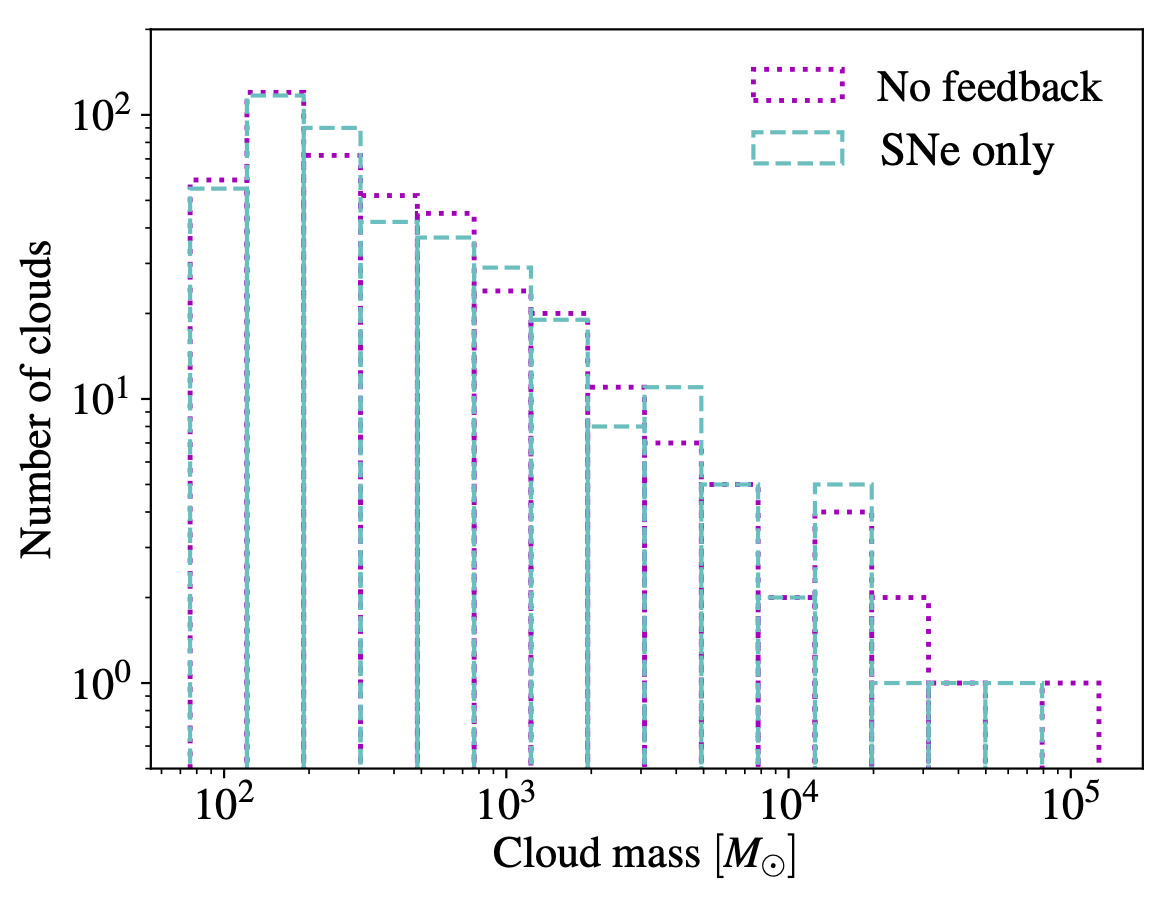}
  \caption[Effect of supernovae on mass of clouds]{Histograms of mass are shown for clouds of the 3 definitions in Section~\ref{sec:clouds}. On the left we compare cloud masses for simulations including both photoionization and SNe, and just photoionization, at a time of 7.5 Myr. On the right we compare the model with SNe with having no feedback. Ionization has the greatest impact on cloud masses, reducing the maximum cloud mass significantly. SNe also reduce cloud masses but have a much smaller impact.}
  \label{fig:cloudmass_sn}
\end{figure*}

\begin{figure*}
 \includegraphics[width=0.45\textwidth]{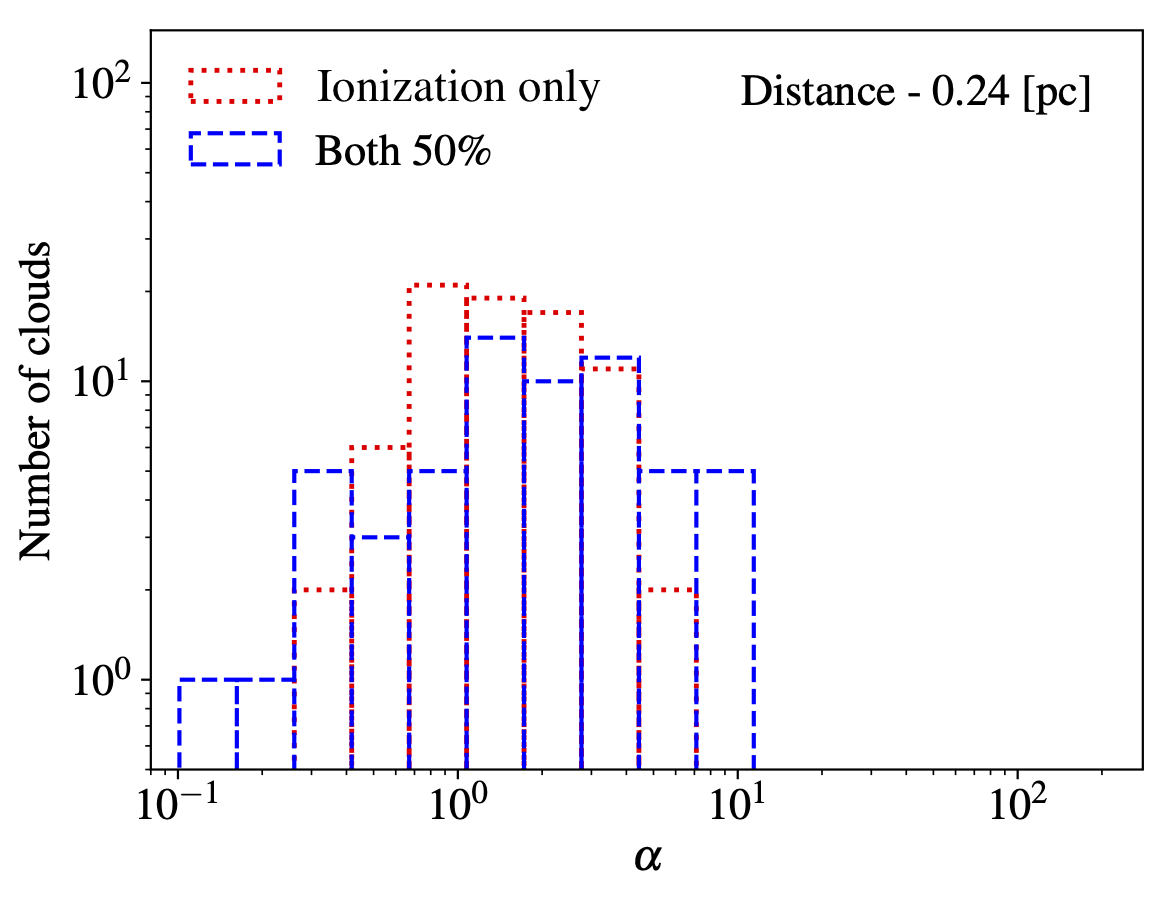}
  \includegraphics[width=0.45\textwidth]{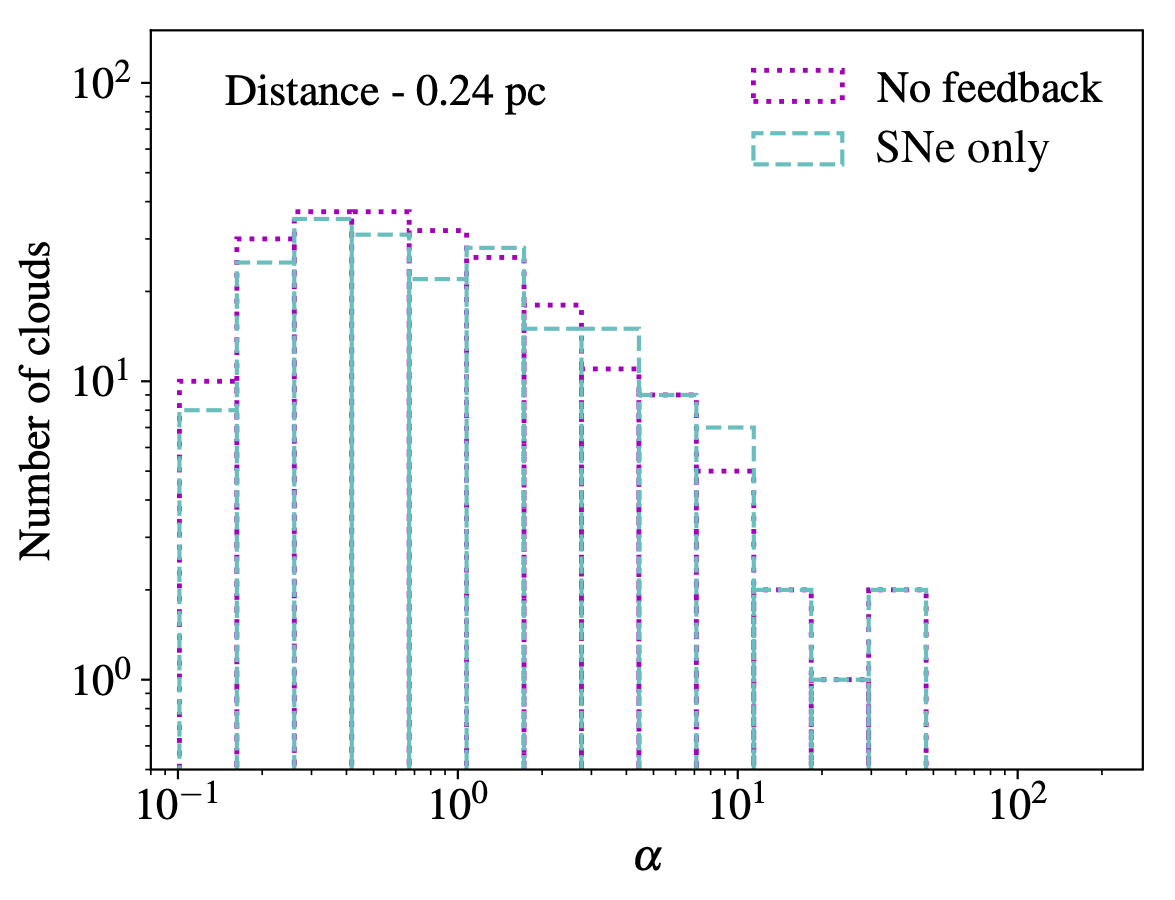}
 \includegraphics[width=0.45\textwidth]{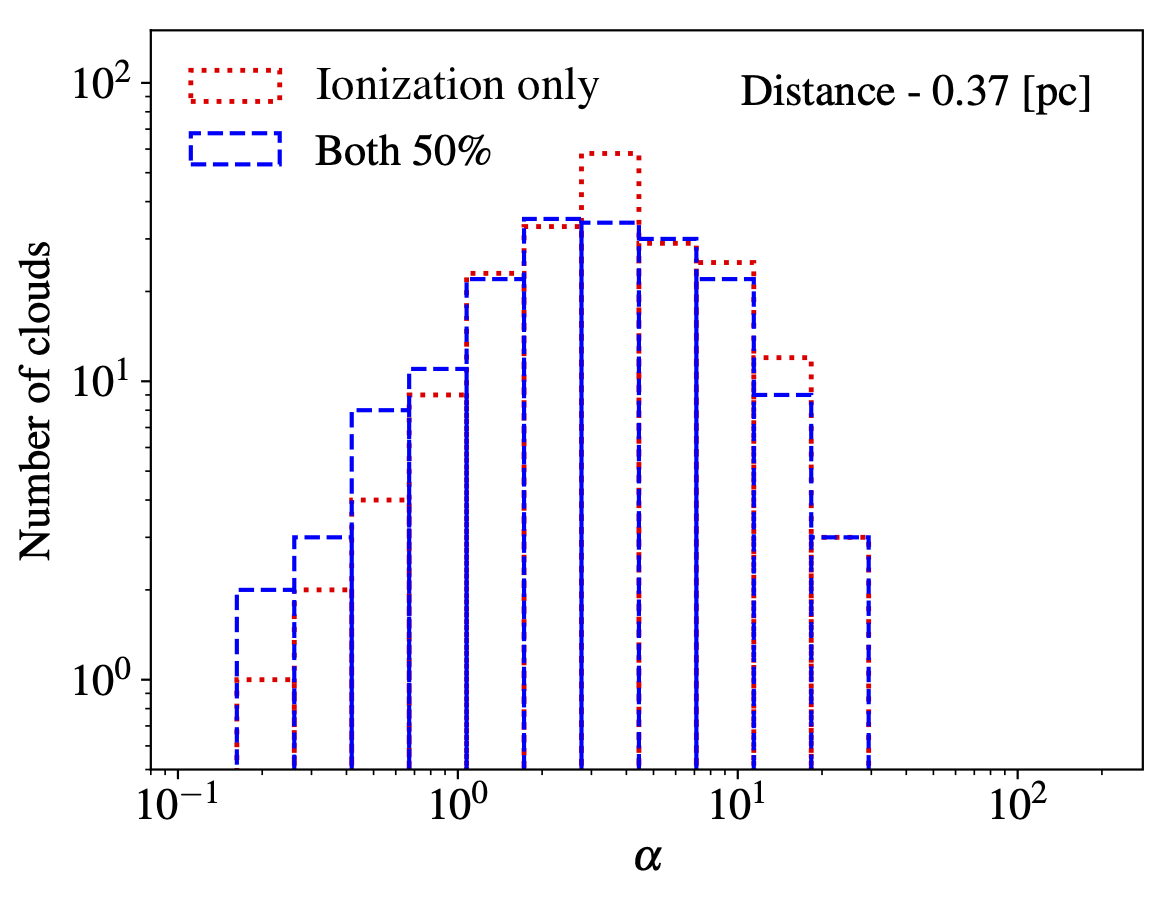}
    \includegraphics[width=0.45\textwidth]{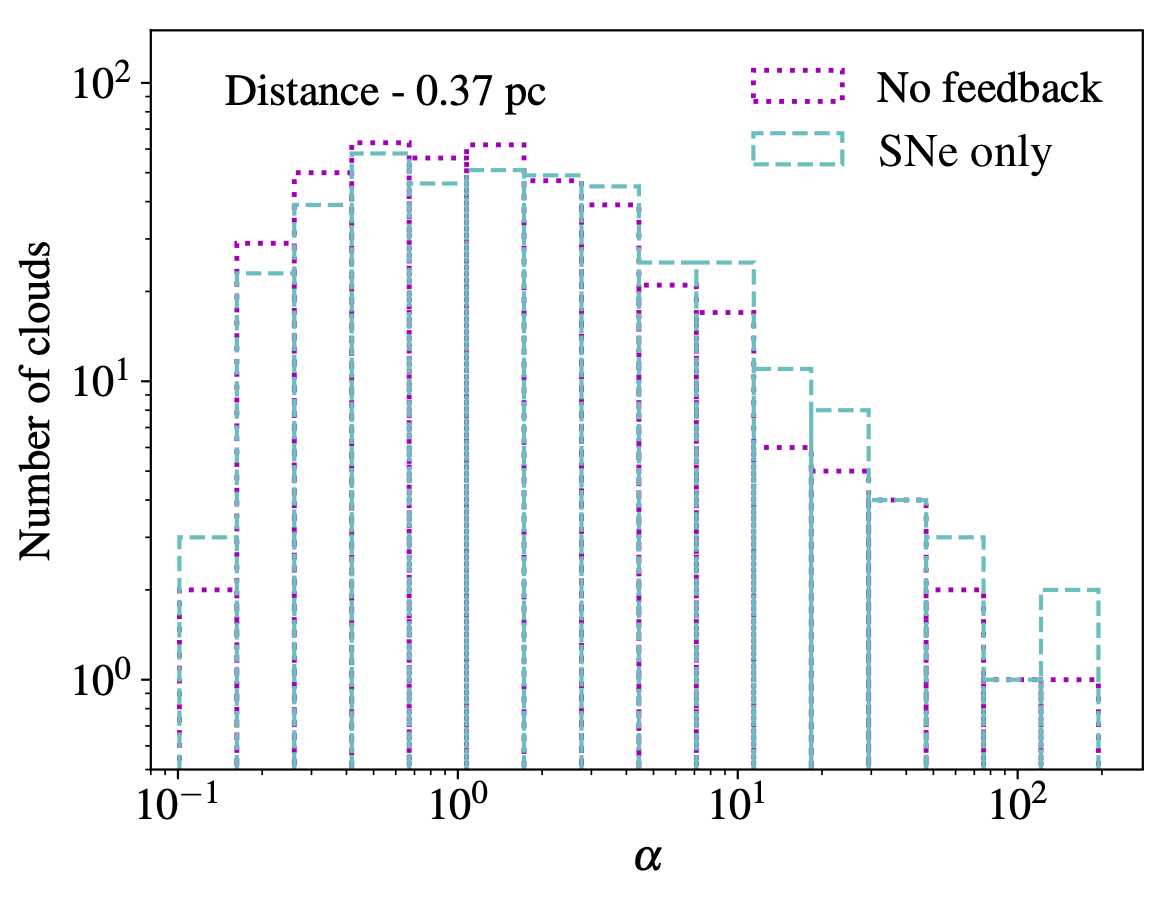}
\includegraphics[width=0.45\textwidth]{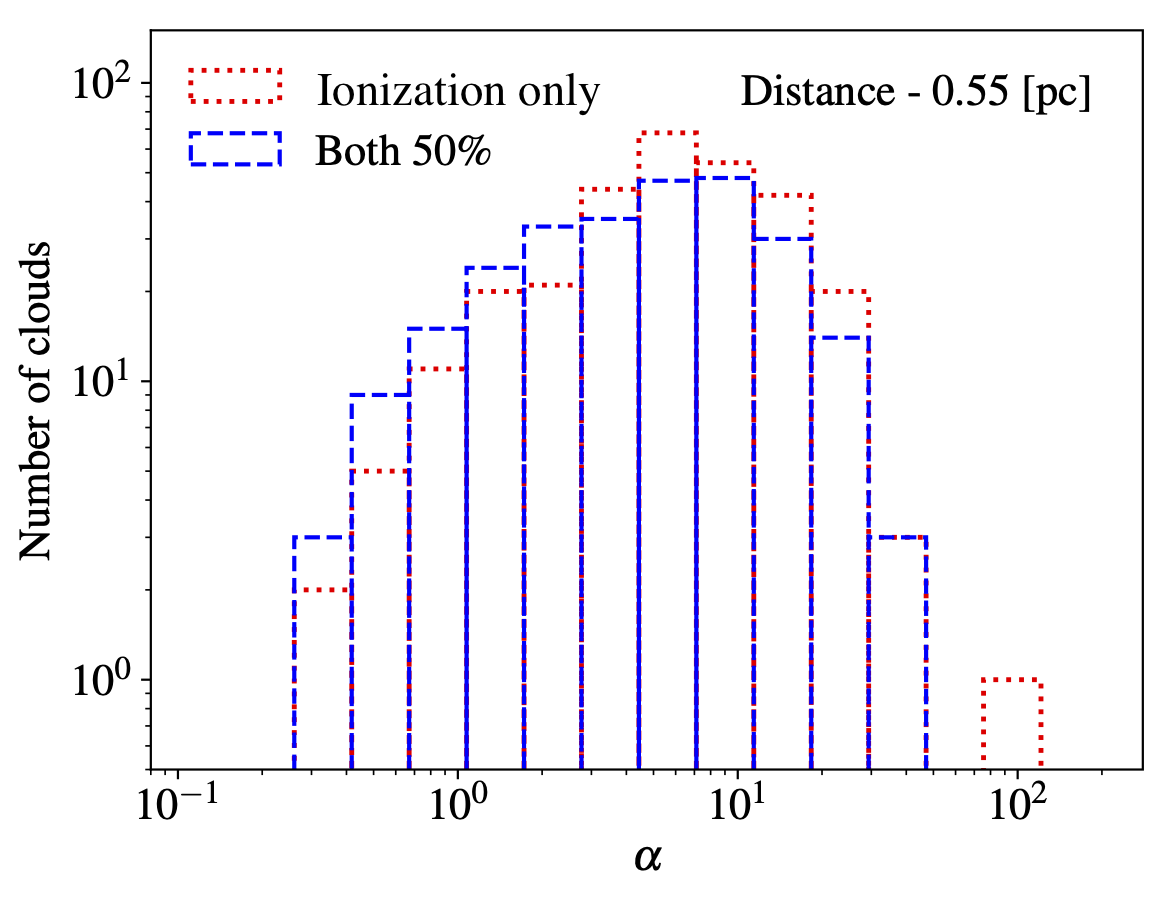}
    \includegraphics[width=0.45\textwidth]{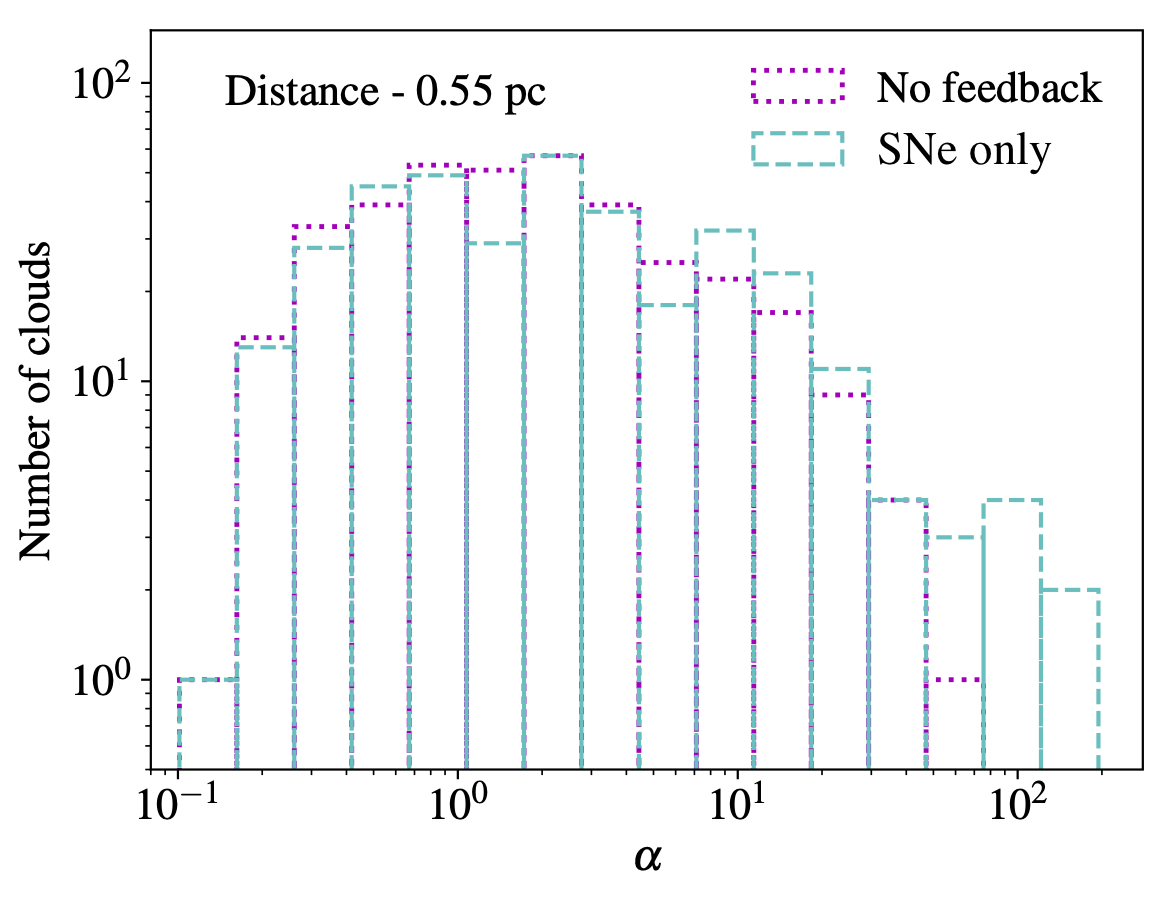}
  \caption[Effect of supernovae on virial parameter of clouds]{Histograms of virial parameter are shown for clouds of the 3 definitions in Section~\ref{sec:clouds}. On the left we compare virial parameters for including both photoionization and SNe, and just photoionization, at a time of 7.5 Myr. On the right we compare the model with SNe with having no feedback. In both cases, ionization has the biggest impact, shifting the peak in the virial parameters to higher values (though reducing the maximum virial parameters). SNe have a little impact, in some cases the number of clouds with high virial parameters is increased slighlty, particularly when there is no ionization.}
    \label{fig:virial_sn}
\end{figure*}

\subsection{Cloud properties} \label{sec:clouds}
In this section we compare the properties of clouds in the four fiducial simulations. 
We find the clouds using the same friends of friends algorithm used in our previous paper \citep{Bending2020}. Particles are considered to be in a cloud if they are within some length scale of at least one other particle in that cloud. We use three distinct length scales 0.24, 0.37 and 0.55 pc, only considering particles above 300, 100 and 25 atoms cm$^{-3}$ respectively. These basically produce features with different surface densities on different scales, those with 0.24 pc are similar to clumps of mass 100-1000 M$_{\odot}$ whilst the length scale of 0.55 pc produces features corresponding to molecular clouds up to $10^4$ or $10^5$ M$_{\odot}$ (see Figure~\ref{fig:cloudmass_sn}).

The impact of photoionization and SNe on cloud masses are shown in Figure~\ref{fig:cloudmass_sn}. Photoionization has the greatest impact on cloud masses (left versus right panels), suppressing the maximum mass by up to an order of magnitude and generally producing fewer clouds except at the lowest masses with the 0.55 pc length scale.  When both forms of feedback are included, supernovae also further lower the numbers of clouds and in particular the number of the most massive clouds, although the effect is much less than photoionization. When photoionization is not included, SNe only have a very slight impact on the cloud masses.

Figure~\ref{fig:virial_sn} shows histograms of the virial parameter of the clouds. 
Similar to the indications from the column density plots and the cloud masses, we find that the biggest differences to the virial parameters occur with (left) and without (right) photoionization. As shown in \citet{Bending2020}, the distribution of $\alpha$ is shifted to higher values of $\alpha$ with photoionization and the minimum and peak in $\alpha$ tend to be higher (though the maximum $\alpha$ is slight lower). Supernovae have relatively little effect on the cloud virial parameters, however they do appear to lead to a higher maximum value of $\alpha$, so a few clouds with larger virial parameter. Generally the supernovae appear to slightly broaden the distribution of $\alpha$. So whilst both the photoionization and SNe appear to increase the velocity dispersion of the gas generally (Section~\ref{sec:veldisp}), photoionization appears to more strongly determine the kinematics of clouds. 

 
\section{Discussion and Conclusions}
\label{sec:conclusion}
We have carried out simulations investigating the combined effect of photoionization and SNe on the ISM along a section of spiral arm. We find that photoionization appears to have a much greater impact than SNe feedback on the structure of the gas. In agreement with previous studies, SNe also have a different impact depending on whether they occur in low density regions, where ionizing feedback has occurred, compared to high density regions, if ionizing feedback is not included. Photoionization converts the surrounding dense gas to warm low density gas. SNe which occur in this environment lead to more spherical shaped bubbles, and simply heat gas which is already warm to even higher temperatures. Any dense gas, and young stars, which lie at the edge of an ionisation region may then appear at the edge of a SNe bubble. 

When SNe occur in the absence of photoionization, they tend to cool more quickly and are less able to heat up the surrounding gas to high temperatures. Also because they are expanding into higher density, structured gas, they tend to form highly asymmetric bubbles. Those that do heat up the gas lead to clearer holes due to the stronger contrast in densities between the supernova remnant and surrounding gas. For our 'Both 50\%' model, photoionizing feedback has a very strong impact on the gas. We tested a less extreme case with our `Both 10\%' model, which includes a lower and likely a more realistic level of photoionization. This still shows photoionizing feedback dominating SNe, and the SNe bubbles forming in low density medium, as described above.

Although we used the same SNe feedback prescription as \citet{Dobbs2011}, where the feedback fairly readily dispersed clouds, the outcome of the feedback on the smaller scales here is somewhat different. This could be due to a number of reasons. Firstly the SNe in these simulations are inserted later (i.e. 5 Myr or more after star formation) which coupled with the higher resolution means in the SNe only run they are inserted into denser gas. Secondly the higher resolution means that the SNe feedback preferentially fills low density regions and channels between high density filaments. These simply would not have been resolved in lower resolution galaxy scale simulations. In these higher resolution simulations, the photoionizing feedback has the strongest impact on dispersing clouds.
We also see with the photoionizing feedback that continuing photoionization past the point SNe occur is incorrect because this produces too much warm gas rather than maintaining a two phase medium. 

It is important to point out that this work does not resolve cluster dynamics and therefore underestimates the influence of runaway O and B stars. \citet{Drew2021} find that between 10 and 20 per cent of O stars in the Carina Arm are runaways. As a result of this the impact of SNe on the ISM may be underestimated in this work.

We also investigate the contribution of photoionization and SNe to turbulent motions in the ISM. We find that photoionization or SNe tend to be a little lower than observed velocity dispersions, and in the ionization only model, the dispersions are starting to decay slightly. Our 'Both 50\%' model produces too high velocities compared with observations, but with a 10\% efficiency, the dispersions agree reasonably well.
We caution though that our simulations are not in equilibrium, and in most models with feedback the velocity dispersions are still fluctuating over the duration of the simulations (and in particular still increasing at larger length scales).

We have not considered either radiation pressure or winds. We leave the first to future work, however in \citet{Ali2022} we found that the impact of winds is less than photoionization and winds act primarily to excavate small $\lesssim 10$ pc cavities around feedback emitting sinks. Likewise the winds contribute to velocity dispersions, but the results from \citet{Ali2022} suggest this occurs on fairly small scales. A second caveat is that we assume some efficiency for the feedback, here our fiducial choice of 50\% was too high and probably 10-20\% is more realistic. This efficiency reflects that we do not resolve down to star-forming scales. The value for the efficiency should correspond to the efficiency at the scales masses of gas are turned into stars each time star formation takes place. These masses in our simulations will be smaller than a molecular cloud but higher than a core, and also dependent on the sink parameters. We further note that at higher resolutions, photoionizing feedback would also be occurring initially in denser gas than we can resolve. Also as mentioned previously, our feedback only occurs in our simulations once stars have formed, and SNe only occur after stars reach the end of their lives, when in reality feedback should be occurring in stars already present at the outset of the simulations.

\section*{Acknowledgements}

Calculations for this paper were performed on the ISCA High Performance Computing Service at the University of Exeter, and used the DiRAC DIaL system, operated by the University of Leicester IT Services, which forms part of the STFC DiRAC HPC Facility (www.dirac.ac.uk).
The equipment was funded by BEIS capital funding via STFC capital grants ST/K000373/1 and ST/R002363/1 and STFC DiRAC Operations grant ST/K001014/1. 
DiRAC is part of the National E-Infrastructure. 
CLD and TJRB acknowledge funding from the European Research Council for the Horizon 2020 ERC consolidator grant project ICYBOB, grant number 818940.

\section*{Data availability}
The data underlying this article will be shared on reasonable request to the corresponding author.




\bibliographystyle{mnras}
\bibliography{library,textbooks,clarebib}

\begin{thebibliography}{}
\makeatletter
\relax
\def\mn@urlcharsother{\let\do\@makeother \do\$\do\&\do\#\do\^\do\_\do\%\do\~}
\def\mn@doi{\begingroup\mn@urlcharsother \@ifnextchar [ {\mn@doi@}
  {\mn@doi@[]}}
\def\mn@doi@[#1]#2{\def\@tempa{#1}\ifx\@tempa\@empty \href
  {http://dx.doi.org/#2} {doi:#2}\else \href {http://dx.doi.org/#2} {#1}\fi
  \endgroup}
\def\mn@eprint#1#2{\mn@eprint@#1:#2::\@nil}
\def\mn@eprint@arXiv#1{\href {http://arxiv.org/abs/#1} {{\tt arXiv:#1}}}
\def\mn@eprint@dblp#1{\href {http://dblp.uni-trier.de/rec/bibtex/#1.xml}
  {dblp:#1}}
\def\mn@eprint@#1:#2:#3:#4\@nil{\def\@tempa {#1}\def\@tempb {#2}\def\@tempc
  {#3}\ifx \@tempc \@empty \let \@tempc \@tempb \let \@tempb \@tempa \fi \ifx
  \@tempb \@empty \def\@tempb {arXiv}\fi \@ifundefined
  {mn@eprint@\@tempb}{\@tempb:\@tempc}{\expandafter \expandafter \csname
  mn@eprint@\@tempb\endcsname \expandafter{\@tempc}}}

\bibitem[\protect\citeauthoryear{Agertz, Kravtsov, Leitner  \& Gnedin}{Agertz
  et~al.}{2013}]{Agertz2013}
Agertz O.,  Kravtsov A.~V.,  Leitner S.~N.,   Gnedin N.~Y.,  2013, \mn@doi
  [ApJ] {10.1088/0004-637X/770/1/25}, 770, 25

\bibitem[\protect\citeauthoryear{Ali, Harries  \& Douglas}{Ali
  et~al.}{2018}]{Ali2018}
Ali A.~A.,  Harries T.~J.,   Douglas T.~A.,  2018, \mn@doi [MNRAS]
  {10.1093/mnras/sty1001}, 477, 5422

\bibitem[\protect\citeauthoryear{{Ali}, {Bending}  \& {Dobbs}}{{Ali}
  et~al.}{2022}]{Ali2022}
{Ali} A.~A.,  {Bending} T. J.~R.,   {Dobbs} C.~L.,  2022, \mn@doi [\mnras]
  {10.1093/mnras/stac025}, \href
  {https://ui.adsabs.harvard.edu/abs/2022MNRAS.510.5592A} {510, 5592}

\bibitem[\protect\citeauthoryear{Bakes \& Tielens}{Bakes \&
  Tielens}{1994}]{Bakes1994}
Bakes E. L.~O.,  Tielens A. G. G.~M.,  1994, ApJ, 427, 822

\bibitem[\protect\citeauthoryear{Bate, Bonnell  \& Price}{Bate
  et~al.}{1995}]{Bate1995}
Bate M.~R.,  Bonnell I.~A.,   Price N.~M.,  1995, MNRAS, 277, 362

\bibitem[\protect\citeauthoryear{Bending, Dobbs  \& Bate}{Bending
  et~al.}{2020}]{Bending2020}
Bending T. J.~R.,  Dobbs C.~L.,   Bate M.~R.,  2020, \mn@doi [MNRAS]
  {10.1093/mnras/staa1293}, 1691, 1672

\bibitem[\protect\citeauthoryear{Benincasa, Wadsley, Couchman, Pettitt, Keller,
  Woods  \& Grond}{Benincasa et~al.}{2020}]{Benincasa2020}
Benincasa S.~M.,  Wadsley J.~W.,  Couchman H.~M.,  Pettitt A.~R.,  Keller
  B.~W.,  Woods R.~M.,   Grond J.~J.,  2020, \mn@doi [MNRAS]
  {10.1093/mnras/staa2935}, 499, 2028

\bibitem[\protect\citeauthoryear{{Benz}}{{Benz}}{1990}]{Benz1990sph-review}
{Benz} W.,  1990, in {Buchler} J.~R.,  ed., Numerical Modelling of Nonlinear
  Stellar Pulsations: Problems and Prospects. Kluwer, Dordrecht, p.~269

\bibitem[\protect\citeauthoryear{Benz, Bowers, Cameron  \& {Press W.H.}}{Benz
  et~al.}{1990}]{Benz1990}
Benz W.,  Bowers R.,  Cameron A.,   {Press W.H.} 1990, ApJ, 348, 647

\bibitem[\protect\citeauthoryear{Binney \& Tremaine}{Binney \&
  Tremaine}{2008}]{binney2008galactic}
Binney J.,  Tremaine S.,  2008, Galactic dynamics.
Princeton University Press, Princeton

\bibitem[\protect\citeauthoryear{{Blum}, {Schaerer}, {Pasquali},
  {Heydari-Malayeri}, {Conti}  \& {Schmutz}}{{Blum} et~al.}{2001}]{Blum2001}
{Blum} R.~D.,  {Schaerer} D.,  {Pasquali} A.,  {Heydari-Malayeri} M.,  {Conti}
  P.~S.,   {Schmutz} W.,  2001, \mn@doi [\aj] {10.1086/323096}, \href
  {https://ui.adsabs.harvard.edu/abs/2001AJ....122.1875B} {122, 1875}

\bibitem[\protect\citeauthoryear{Brunt}{Brunt}{2003}]{Brunt2003}
Brunt C.~M.,  2003, \mn@doi [ApJ] {10.1086/345597}, 584, 293

\bibitem[\protect\citeauthoryear{Butler, Tan, Teyssier, Rosdahl, {Van Loo}  \&
  Nickerson}{Butler et~al.}{2017}]{Butler2017}
Butler M.~J.,  Tan J.~C.,  Teyssier R.,  Rosdahl J.,  {Van Loo} S.,   Nickerson
  S.,  2017, \mn@doi [ApJ] {10.3847/1538-4357/aa7054}, 841, 82

\bibitem[\protect\citeauthoryear{{Calzetti}, {Meurer}, {Bohlin}, {Garnett},
  {Kinney}, {Leitherer}  \& {Storchi-Bergmann}}{{Calzetti}
  et~al.}{1997}]{Calzetti1997}
{Calzetti} D.,  {Meurer} G.~R.,  {Bohlin} R.~C.,  {Garnett} D.~R.,  {Kinney}
  A.~L.,  {Leitherer} C.,   {Storchi-Bergmann} T.,  1997, \mn@doi [\aj]
  {10.1086/118609}, \href
  {https://ui.adsabs.harvard.edu/abs/1997AJ....114.1834C} {114, 1834}

\bibitem[\protect\citeauthoryear{Chevance et~al.,}{Chevance
  et~al.}{2020}]{Chevance2020}
Chevance M.,  et~al., 2020, \mn@doi [MNRAS] {10.1093/mnras/stz3525}, 493, 2872

\bibitem[\protect\citeauthoryear{{Chevance} et~al.,}{{Chevance}
  et~al.}{2022}]{Chevance2022}
{Chevance} M.,  et~al., 2022, \mn@doi [\mnras] {10.1093/mnras/stab2938}, \href
  {https://ui.adsabs.harvard.edu/abs/2022MNRAS.509..272C} {509, 272}

\bibitem[\protect\citeauthoryear{Cioffi, Mckee  \& Bertschinger}{Cioffi
  et~al.}{1988}]{Cioffi1988}
Cioffi D.~F.,  Mckee C.~F.,   Bertschinger E.,  1988, J. Chem. Inf. Model.,
  334, 252

\bibitem[\protect\citeauthoryear{Col{\'{i}}n, V{\'{a}}zquez-Semadeni  \&
  G{\'{o}}mez}{Col{\'{i}}n et~al.}{2013}]{Colin2013}
Col{\'{i}}n P.,  V{\'{a}}zquez-Semadeni E.,   G{\'{o}}mez G.~C.,  2013, \mn@doi
  [MNRAS] {10.1093/mnras/stt1409}, 435, 1701

\bibitem[\protect\citeauthoryear{{Colling}, {Hennebelle}, {Geen}, {Iffrig}  \&
  {Bournaud}}{{Colling} et~al.}{2018}]{Colling2018}
{Colling} C.,  {Hennebelle} P.,  {Geen} S.,  {Iffrig} O.,   {Bournaud} F.,
  2018, \mn@doi [\aap] {10.1051/0004-6361/201833161}, \href
  {https://ui.adsabs.harvard.edu/abs/2018A&A...620A..21C} {620, A21}

\bibitem[\protect\citeauthoryear{Cox \& Gomez}{Cox \& Gomez}{2002}]{Cox2002}
Cox D.~P.,  Gomez G.~C.,  2002, \mn@doi [ApJS] {10.1086/341946}, 142, 261

\bibitem[\protect\citeauthoryear{Dale, Ercolano  \& Bonnell}{Dale
  et~al.}{2012}]{Dale2012a}
Dale J.~E.,  Ercolano B.,   Bonnell I.~A.,  2012, \mn@doi [MNRAS]
  {10.1111/j.1365-2966.2012.21205.x}, 424, 377

\bibitem[\protect\citeauthoryear{Dale, Ercolano  \& Bonnell}{Dale
  et~al.}{2013}]{Dale2013}
Dale J.~E.,  Ercolano B.,   Bonnell I.~A.,  2013, \mn@doi [MNRAS]
  {10.1093/mnras/sts592}, 430, 234

\bibitem[\protect\citeauthoryear{{Dalla Vecchia} \& {Schaye}}{{Dalla Vecchia}
  \& {Schaye}}{2012}]{Dalla2012}
{Dalla Vecchia} C.,  {Schaye} J.,  2012, \mn@doi [\mnras]
  {10.1111/j.1365-2966.2012.21704.x}, \href
  {https://ui.adsabs.harvard.edu/abs/2012MNRAS.426..140D} {426, 140}

\bibitem[\protect\citeauthoryear{Dobbs \& Pringle}{Dobbs \&
  Pringle}{2013}]{Dobbs2013}
Dobbs C.~L.,  Pringle J.~E.,  2013, \mn@doi [MNRAS] {10.1093/mnras/stt508},
  432, 653

\bibitem[\protect\citeauthoryear{Dobbs, Glover, Clark  \& Klessen}{Dobbs
  et~al.}{2008}]{Dobbs2008}
Dobbs C.~L.,  Glover S. C.~O.,  Clark P.~C.,   Klessen R.~S.,  2008, \mn@doi
  [MNRAS] {10.1111/j.1365-2966.2008.13646.x}, 389, 1097

\bibitem[\protect\citeauthoryear{Dobbs, Burkert  \& Pringle}{Dobbs
  et~al.}{2011}]{Dobbs2011}
Dobbs C.~L.,  Burkert A.,   Pringle J.~E.,  2011, \mn@doi [MNRAS]
  {10.1111/j.1365-2966.2011.19346.x}, 417, 1318

\bibitem[\protect\citeauthoryear{Drew, Mongui  \& Wright}{Drew
  et~al.}{2021}]{Drew2021}
Drew J.~E.,  Mongui M.,   Wright N.~J.,  2021, MNRAS, 508, 4952

\bibitem[\protect\citeauthoryear{{Efstathiou}}{{Efstathiou}}{2000}]{Efstathiou2000}
{Efstathiou} G.,  2000, \mn@doi [\mnras] {10.1046/j.1365-8711.2000.03665.x},
  \href {https://ui.adsabs.harvard.edu/abs/2000MNRAS.317..697E} {317, 697}

\bibitem[\protect\citeauthoryear{{Elmegreen} \& {Scalo}}{{Elmegreen} \&
  {Scalo}}{2004}]{Elmegreen2004}
{Elmegreen} B.~G.,  {Scalo} J.,  2004, \mn@doi [\araa]
  {10.1146/annurev.astro.41.011802.094859}, \href
  {https://ui.adsabs.harvard.edu/abs/2004ARA&A..42..211E} {42, 211}

\bibitem[\protect\citeauthoryear{Emerick, Bryan  \& Low}{Emerick
  et~al.}{2018}]{Emerick2018}
Emerick A.,  Bryan G.~L.,   Low M.-M.~M.,  2018, \mn@doi [Astrophys. J. Lett.]
  {10.3847/2041-8213/aae315}, 865, L22

\bibitem[\protect\citeauthoryear{Fukushima, Yajima, Sugimura, Hosokawa, Omukai
  \& Matsumoto}{Fukushima et~al.}{2020}]{Fukushima2020a}
Fukushima H.,  Yajima H.,  Sugimura K.,  Hosokawa T.,  Omukai K.,   Matsumoto
  T.,  2020, \mn@doi [MNRAS] {10.1093/mnras/staa2062}, 497, 3830

\bibitem[\protect\citeauthoryear{Geen, Hennebelle, Tremblin  \& Rosdahl}{Geen
  et~al.}{2016}]{Geen2016}
Geen S.,  Hennebelle P.,  Tremblin P.,   Rosdahl J.,  2016, \mn@doi [MNRAS]
  {10.1093/mnras/stw2235}, 463, 3129

\bibitem[\protect\citeauthoryear{Geen, Watson, Rosdahl, Bieri, Klessen  \&
  Hennebelle}{Geen et~al.}{2018}]{Geen2018}
Geen S.,  Watson S.~K.,  Rosdahl J.,  Bieri R.,  Klessen R.~S.,   Hennebelle
  P.,  2018, MNRAS, 481, 2548

\bibitem[\protect\citeauthoryear{{Gentry}, {Krumholz}, {Dekel}  \&
  {Madau}}{{Gentry} et~al.}{2017}]{Gentry2017}
{Gentry} E.~S.,  {Krumholz} M.~R.,  {Dekel} A.,   {Madau} P.,  2017, \mn@doi
  [\mnras] {10.1093/mnras/stw2746}, \href
  {https://ui.adsabs.harvard.edu/abs/2017MNRAS.465.2471G} {465, 2471}

\bibitem[\protect\citeauthoryear{{Gerritsen} \& {Icke}}{{Gerritsen} \&
  {Icke}}{1997}]{Gerritsen1997}
{Gerritsen} J.~P.~E.,  {Icke} V.,  1997, \aap, \href
  {https://ui.adsabs.harvard.edu/abs/1997A&A...325..972G} {325, 972}

\bibitem[\protect\citeauthoryear{Glover \& {Mac Low}}{Glover \& {Mac
  Low}}{2007}]{Glover2007}
Glover S. C.~O.,  {Mac Low} M.,  2007, \mn@doi [ApJS] {10.1086/512238}, 169,
  239

\bibitem[\protect\citeauthoryear{{Governato}, {Willman}, {Mayer}, {Brooks},
  {Stinson}, {Valenzuela}, {Wadsley}  \& {Quinn}}{{Governato}
  et~al.}{2007}]{Governato2007}
{Governato} F.,  {Willman} B.,  {Mayer} L.,  {Brooks} A.,  {Stinson} G.,
  {Valenzuela} O.,  {Wadsley} J.,   {Quinn} T.,  2007, \mn@doi [\mnras]
  {10.1111/j.1365-2966.2006.11266.x}, \href
  {https://ui.adsabs.harvard.edu/abs/2007MNRAS.374.1479G} {374, 1479}

\bibitem[\protect\citeauthoryear{Grasha et~al.,}{Grasha
  et~al.}{2019}]{Grasha2019}
Grasha K.,  et~al., 2019, \mn@doi [MNRAS] {10.1093/mnras/sty3424}, 483, 4707

\bibitem[\protect\citeauthoryear{Grisdale, Agertz, Renaud  \& Romeo}{Grisdale
  et~al.}{2018}]{Grisdale2018}
Grisdale K.,  Agertz O.,  Renaud F.,   Romeo A.~B.,  2018, \mn@doi [MNRAS]
  {10.1093/mnras/sty1595}, 479, 3167

\bibitem[\protect\citeauthoryear{{Grudi{\'c}}, {Hopkins},
  {Faucher-Gigu{\`e}re}, {Quataert}, {Murray}  \& {Kere{\v{s}}}}{{Grudi{\'c}}
  et~al.}{2018}]{Grudic2018}
{Grudi{\'c}} M.~Y.,  {Hopkins} P.~F.,  {Faucher-Gigu{\`e}re} C.-A.,  {Quataert}
  E.,  {Murray} N.,   {Kere{\v{s}}} D.,  2018, \mn@doi [\mnras]
  {10.1093/mnras/sty035}, \href
  {https://ui.adsabs.harvard.edu/abs/2018MNRAS.475.3511G} {475, 3511}

\bibitem[\protect\citeauthoryear{Habing}{Habing}{1968}]{Habing1968}
Habing H.~J.,  1968, Bull. Astron. Institutes Netherlands, 19, 421

\bibitem[\protect\citeauthoryear{{Haehnelt}}{{Haehnelt}}{1995}]{Haehnelt1995}
{Haehnelt} M.~G.,  1995, \mn@doi [\mnras] {10.1093/mnras/273.2.249}, \href
  {https://ui.adsabs.harvard.edu/abs/1995MNRAS.273..249H} {273, 249}

\bibitem[\protect\citeauthoryear{{Haid}, {Walch}, {Seifried}, {W{\"u}nsch},
  {Dinnbier}  \& {Naab}}{{Haid} et~al.}{2019}]{Haid2019}
{Haid} S.,  {Walch} S.,  {Seifried} D.,  {W{\"u}nsch} R.,  {Dinnbier} F.,
  {Naab} T.,  2019, \mn@doi [\mnras] {10.1093/mnras/sty2938}, \href
  {https://ui.adsabs.harvard.edu/abs/2019MNRAS.482.4062H} {482, 4062}

\bibitem[\protect\citeauthoryear{Heyer \& Brunt}{Heyer \&
  Brunt}{2004}]{Heyer2004}
Heyer M.~H.,  Brunt C.~M.,  2004, \mn@doi [ApJ] {10.1086/425978}, 615, L45

\bibitem[\protect\citeauthoryear{Hill, {Ryan Joung}, {Mac Low}, Benjamin,
  {Matthew Haffner}, Klingenberg  \& Waagan}{Hill et~al.}{2012}]{Hill2012}
Hill A.~S.,  {Ryan Joung} M.,  {Mac Low} M.~M.,  Benjamin R.~A.,  {Matthew
  Haffner} L.,  Klingenberg C.,   Waagan K.,  2012, \mn@doi [ApJ]
  {10.1088/0004-637X/750/2/104}, 750, 104

\bibitem[\protect\citeauthoryear{{Hollyhead}, {Bastian}, {Adamo},
  {Silva-Villa}, {Dale}, {Ryon}  \& {Gazak}}{{Hollyhead}
  et~al.}{2015}]{Hollyhead2015}
{Hollyhead} K.,  {Bastian} N.,  {Adamo} A.,  {Silva-Villa} E.,  {Dale} J.,
  {Ryon} J.~E.,   {Gazak} Z.,  2015, \mn@doi [\mnras] {10.1093/mnras/stv331},
  \href {https://ui.adsabs.harvard.edu/abs/2015MNRAS.449.1106H} {449, 1106}

\bibitem[\protect\citeauthoryear{{Hopkins} et~al.,}{{Hopkins}
  et~al.}{2018a}]{Hopkins2018}
{Hopkins} P.~F.,  et~al., 2018a, \mn@doi [\mnras] {10.1093/mnras/sty674}, \href
  {https://ui.adsabs.harvard.edu/abs/2018MNRAS.477.1578H} {477, 1578}

\bibitem[\protect\citeauthoryear{{Hopkins} et~al.,}{{Hopkins}
  et~al.}{2018b}]{Hopkins2018a}
{Hopkins} P.~F.,  et~al., 2018b, \mn@doi [\mnras] {10.1093/mnras/sty1690},
  \href {https://ui.adsabs.harvard.edu/abs/2018MNRAS.480..800H} {480, 800}

\bibitem[\protect\citeauthoryear{{Hu}, {Naab}, {Walch}, {Glover}  \&
  {Clark}}{{Hu} et~al.}{2016}]{Hu2016}
{Hu} C.-Y.,  {Naab} T.,  {Walch} S.,  {Glover} S. C.~O.,   {Clark} P.~C.,
  2016, \mn@doi [\mnras] {10.1093/mnras/stw544}, \href
  {https://ui.adsabs.harvard.edu/abs/2016MNRAS.458.3528H} {458, 3528}

\bibitem[\protect\citeauthoryear{Ikeuchi, Habe  \& Tanaka}{Ikeuchi
  et~al.}{1984}]{Ikeuchi1984}
Ikeuchi S.,  Habe A.,   Tanaka Y.~D.,  1984, MNRAS, 207, 909

\bibitem[\protect\citeauthoryear{{Jeffreson}, {Kruijssen}, {Keller}, {Chevance}
   \& {Glover}}{{Jeffreson} et~al.}{2020}]{Jeffreson2020}
{Jeffreson} S. M.~R.,  {Kruijssen} J.~M.~D.,  {Keller} B.~W.,  {Chevance} M.,
  {Glover} S. C.~O.,  2020, \mn@doi [\mnras] {10.1093/mnras/staa2127}, \href
  {https://ui.adsabs.harvard.edu/abs/2020MNRAS.498..385J} {498, 385}

\bibitem[\protect\citeauthoryear{{Jeffreson}, {Krumholz}, {Fujimoto},
  {Armillotta}, {Keller}, {Chevance}  \& {Kruijssen}}{{Jeffreson}
  et~al.}{2021}]{Jeffreson2021}
{Jeffreson} S. M.~R.,  {Krumholz} M.~R.,  {Fujimoto} Y.,  {Armillotta} L.,
  {Keller} B.~W.,  {Chevance} M.,   {Kruijssen} J.~M.~D.,  2021, \mn@doi
  [\mnras] {10.1093/mnras/stab1536}, \href
  {https://ui.adsabs.harvard.edu/abs/2021MNRAS.505.3470J} {505, 3470}

\bibitem[\protect\citeauthoryear{{Joung} \& {Mac Low}}{{Joung} \& {Mac
  Low}}{2006}]{Joung2006}
{Joung} M.~K.~R.,  {Mac Low} M.-M.,  2006, \mn@doi [\apj] {10.1086/508795},
  \href {https://ui.adsabs.harvard.edu/abs/2006ApJ...653.1266J} {653, 1266}

\bibitem[\protect\citeauthoryear{{Kannan}, {Marinacci}, {Simpson}, {Glover}  \&
  {Hernquist}}{{Kannan} et~al.}{2020}]{Kannan2020}
{Kannan} R.,  {Marinacci} F.,  {Simpson} C.~M.,  {Glover} S. C.~O.,
  {Hernquist} L.,  2020, \mn@doi [\mnras] {10.1093/mnras/stz3078}, \href
  {https://ui.adsabs.harvard.edu/abs/2020MNRAS.491.2088K} {491, 2088}

\bibitem[\protect\citeauthoryear{{Katz}}{{Katz}}{1992}]{Katz1992}
{Katz} N.,  1992, \mn@doi [\apj] {10.1086/171366}, \href
  {https://ui.adsabs.harvard.edu/abs/1992ApJ...391..502K} {391, 502}

\bibitem[\protect\citeauthoryear{{Kay}, {Pearce}, {Frenk}  \& {Jenkins}}{{Kay}
  et~al.}{2002}]{Kay2002}
{Kay} S.~T.,  {Pearce} F.~R.,  {Frenk} C.~S.,   {Jenkins} A.,  2002, \mn@doi
  [\mnras] {10.1046/j.1365-8711.2002.05070.x}, \href
  {https://ui.adsabs.harvard.edu/abs/2002MNRAS.330..113K} {330, 113}

\bibitem[\protect\citeauthoryear{{Keller}, {Wadsley}, {Benincasa}  \&
  {Couchman}}{{Keller} et~al.}{2014}]{Keller2014}
{Keller} B.~W.,  {Wadsley} J.,  {Benincasa} S.~M.,   {Couchman} H.~M.~P.,
  2014, \mn@doi [\mnras] {10.1093/mnras/stu1058}, \href
  {https://ui.adsabs.harvard.edu/abs/2014MNRAS.442.3013K} {442, 3013}

\bibitem[\protect\citeauthoryear{{Kim} \& {Ostriker}}{{Kim} \&
  {Ostriker}}{2017}]{Kim2017a}
{Kim} C.-G.,  {Ostriker} E.~C.,  2017, \mn@doi [\apj]
  {10.3847/1538-4357/aa8599}, \href
  {https://ui.adsabs.harvard.edu/abs/2017ApJ...846..133K} {846, 133}

\bibitem[\protect\citeauthoryear{{Kim}, {Ostriker}  \& {Raileanu}}{{Kim}
  et~al.}{2017}]{Kim2017}
{Kim} C.-G.,  {Ostriker} E.~C.,   {Raileanu} R.,  2017, \mn@doi [\apj]
  {10.3847/1538-4357/834/1/25}, \href
  {https://ui.adsabs.harvard.edu/abs/2017ApJ...834...25K} {834, 25}

\bibitem[\protect\citeauthoryear{Kim, Kim  \& Ostriker}{Kim
  et~al.}{2018}]{Kim2018}
Kim J.-G.,  Kim W.-T.,   Ostriker E.~C.,  2018, \mn@doi [ApJ]
  {10.3847/1538-4357/aabe27}, 859, 68

\bibitem[\protect\citeauthoryear{{Kimm}, {Cen}, {Devriendt}, {Dubois}  \&
  {Slyz}}{{Kimm} et~al.}{2015}]{Kimm2015}
{Kimm} T.,  {Cen} R.,  {Devriendt} J.,  {Dubois} Y.,   {Slyz} A.,  2015,
  \mn@doi [\mnras] {10.1093/mnras/stv1211}, \href
  {https://ui.adsabs.harvard.edu/abs/2015MNRAS.451.2900K} {451, 2900}

\bibitem[\protect\citeauthoryear{{Kravtsov} \& {Yepes}}{{Kravtsov} \&
  {Yepes}}{2000}]{Kravtsov2000}
{Kravtsov} A.~V.,  {Yepes} G.,  2000, \mn@doi [\mnras]
  {10.1046/j.1365-8711.2000.03771.x}, \href
  {https://ui.adsabs.harvard.edu/abs/2000MNRAS.318..227K} {318, 227}

\bibitem[\protect\citeauthoryear{Kroupa}{Kroupa}{2001}]{Kroupa2001}
Kroupa P.,  2001, \mn@doi [MNRAS] {10.1046/j.1365-8711.2001.04022.x}, 246, 231

\bibitem[\protect\citeauthoryear{Larson}{Larson}{1981}]{Larson1981}
Larson R.~B.,  1981, MNRAS, 194, 809

\bibitem[\protect\citeauthoryear{{Lu}, {Pelkonen}, {Padoan}, {Pan},
  {Haugb{\o}lle}  \& {Nordlund}}{{Lu} et~al.}{2020}]{Lu2020}
{Lu} Z.-J.,  {Pelkonen} V.-M.,  {Padoan} P.,  {Pan} L.,  {Haugb{\o}lle} T.,
  {Nordlund} {\r{A}}.,  2020, \mn@doi [\apj] {10.3847/1538-4357/abbd8f}, \href
  {https://ui.adsabs.harvard.edu/abs/2020ApJ...904...58L} {904, 58}

\bibitem[\protect\citeauthoryear{Lucas, Bonnell  \& Dale}{Lucas
  et~al.}{2020}]{Lucas2020}
Lucas W.~E.,  Bonnell I.~A.,   Dale J.~E.,  2020, \mn@doi [MNRAS]
  {10.1093/mnras/staa451}, 493, 4700

\bibitem[\protect\citeauthoryear{{Mac Low} \& {Klessen}}{{Mac Low} \&
  {Klessen}}{2004}]{MacLow2004}
{Mac Low} M.-M.,  {Klessen} R.~S.,  2004, \mn@doi [Reviews of Modern Physics]
  {10.1103/RevModPhys.76.125}, \href
  {https://ui.adsabs.harvard.edu/abs/2004RvMP...76..125M} {76, 125}

\bibitem[\protect\citeauthoryear{{McLeod} et~al.,}{{McLeod}
  et~al.}{2021}]{McLeod2021}
{McLeod} A.~F.,  et~al., 2021, \mn@doi [\mnras] {10.1093/mnras/stab2726}, \href
  {https://ui.adsabs.harvard.edu/abs/2021MNRAS.508.5425M} {508, 5425}

\bibitem[\protect\citeauthoryear{{Navarro} \& {White}}{{Navarro} \&
  {White}}{1993}]{Navarro1993}
{Navarro} J.~F.,  {White} S.~D.~M.,  1993, \mn@doi [\mnras]
  {10.1093/mnras/265.2.271}, \href
  {https://ui.adsabs.harvard.edu/abs/1993MNRAS.265..271N} {265, 271}

\bibitem[\protect\citeauthoryear{{Nguyen-Luong} et~al.,}{{Nguyen-Luong}
  et~al.}{2016}]{Luong2016}
{Nguyen-Luong} Q.,  et~al., 2016, \mn@doi [\apj] {10.3847/0004-637X/833/1/23},
  \href {https://ui.adsabs.harvard.edu/abs/2016ApJ...833...23N} {833, 23}

\bibitem[\protect\citeauthoryear{Peters et~al.,}{Peters
  et~al.}{2017}]{Peters2017}
Peters T.,  et~al., 2017, \mn@doi [MNRAS] {10.1093/mnras/stw3216}, 466, 3293

\bibitem[\protect\citeauthoryear{{Portegies Zwart} \& {Verbunt}}{{Portegies
  Zwart} \& {Verbunt}}{2012}]{Portegies2012}
{Portegies Zwart} S.~F.,  {Verbunt} F.,  2012, {SeBa: Stellar and binary
  evolution} (\mn@eprint {ascl} {1201.003})

\bibitem[\protect\citeauthoryear{Price}{Price}{2007}]{Price2007}
Price D.~J.,  2007, \mn@doi [Publ. Astron. Soc. Aust.] {10.1071/AS07022}, 24,
  159

\bibitem[\protect\citeauthoryear{{Rathjen} et~al.,}{{Rathjen}
  et~al.}{2021}]{Rathjen2021}
{Rathjen} T.-E.,  et~al., 2021, \mn@doi [\mnras] {10.1093/mnras/stab900}, \href
  {https://ui.adsabs.harvard.edu/abs/2021MNRAS.504.1039R} {504, 1039}

\bibitem[\protect\citeauthoryear{Rogers \& Pittard}{Rogers \&
  Pittard}{2013}]{Rogers2013}
Rogers H.,  Pittard J.~M.,  2013, \mn@doi [MNRAS] {10.1093/mnras/stu625}, 431,
  1337

\bibitem[\protect\citeauthoryear{Saitoh \& Makino}{Saitoh \&
  Makino}{2009}]{Saitoh2009}
Saitoh T.~R.,  Makino J.,  2009, \mn@doi [ApJ] {10.1088/0004-637X/697/2/L99},
  697, L99

\bibitem[\protect\citeauthoryear{{Scannapieco}, {Tissera}, {White}  \&
  {Springel}}{{Scannapieco} et~al.}{2006}]{Scan2006}
{Scannapieco} C.,  {Tissera} P.~B.,  {White} S.~D.~M.,   {Springel} V.,  2006,
  \mn@doi [\mnras] {10.1111/j.1365-2966.2006.10785.x}, \href
  {https://ui.adsabs.harvard.edu/abs/2006MNRAS.371.1125S} {371, 1125}

\bibitem[\protect\citeauthoryear{Seifried, Walch, Haid, Girichidis  \&
  Naab}{Seifried et~al.}{2018}]{Seifried2018}
Seifried D.,  Walch S.,  Haid S.,  Girichidis P.,   Naab T.,  2018, \mn@doi
  [ApJ] {10.3847/1538-4357/aaacff}, 855, 81

\bibitem[\protect\citeauthoryear{Solomon, Rivolo, Barrett  \& Yahil}{Solomon
  et~al.}{1987}]{Solomon1987}
Solomon P.~M.,  Rivolo A.~R.,  Barrett J.,   Yahil A.,  1987, ApJ, 319, 730

\bibitem[\protect\citeauthoryear{{Sommer-Larsen}, {Gelato}  \&
  {Vedel}}{{Sommer-Larsen} et~al.}{1999}]{Sommer1999}
{Sommer-Larsen} J.,  {Gelato} S.,   {Vedel} H.,  1999, \mn@doi [\apj]
  {10.1086/307374}, \href
  {https://ui.adsabs.harvard.edu/abs/1999ApJ...519..501S} {519, 501}

\bibitem[\protect\citeauthoryear{{Springel}}{{Springel}}{2000}]{Springel2000}
{Springel} V.,  2000, \mn@doi [\mnras] {10.1046/j.1365-8711.2000.03187.x},
  \href {https://ui.adsabs.harvard.edu/abs/2000MNRAS.312..859S} {312, 859}

\bibitem[\protect\citeauthoryear{{Stinson}, {Seth}, {Katz}, {Wadsley},
  {Governato}  \& {Quinn}}{{Stinson} et~al.}{2006}]{Stinson2006}
{Stinson} G.,  {Seth} A.,  {Katz} N.,  {Wadsley} J.,  {Governato} F.,   {Quinn}
  T.,  2006, \mn@doi [\mnras] {10.1111/j.1365-2966.2006.11097.x}, \href
  {https://ui.adsabs.harvard.edu/abs/2006MNRAS.373.1074S} {373, 1074}

\bibitem[\protect\citeauthoryear{{Stinson} et~al.,}{{Stinson}
  et~al.}{2013}]{Stinson2013}
{Stinson} G.~S.,  et~al., 2013, \mn@doi [\mnras] {10.1093/mnras/stt1600}, \href
  {https://ui.adsabs.harvard.edu/abs/2013MNRAS.436..625S} {436, 625}

\bibitem[\protect\citeauthoryear{{Thacker} \& {Couchman}}{{Thacker} \&
  {Couchman}}{2000}]{Thacker2000}
{Thacker} R.~J.,  {Couchman} H.~M.~P.,  2000, \mn@doi [\apj] {10.1086/317828},
  \href {https://ui.adsabs.harvard.edu/abs/2000ApJ...545..728T} {545, 728}

\bibitem[\protect\citeauthoryear{{Thacker} \& {Couchman}}{{Thacker} \&
  {Couchman}}{2001}]{Thacker2001}
{Thacker} R.~J.,  {Couchman} H.~M.~P.,  2001, \mn@doi [\apjl] {10.1086/321739},
  \href {https://ui.adsabs.harvard.edu/abs/2001ApJ...555L..17T} {555, L17}

\bibitem[\protect\citeauthoryear{{Thornton}, {Gaudlitz}, {Janka}  \&
  {Steinmetz}}{{Thornton} et~al.}{1998}]{Thornton1998}
{Thornton} K.,  {Gaudlitz} M.,  {Janka} H.~T.,   {Steinmetz} M.,  1998, \mn@doi
  [\apj] {10.1086/305704}, \href
  {https://ui.adsabs.harvard.edu/abs/1998ApJ...500...95T} {500, 95}

\bibitem[\protect\citeauthoryear{Toonen, Nelemans  \& {Portegies Zwart}}{Toonen
  et~al.}{2012}]{Toonen2012}
Toonen S.,  Nelemans G.,   {Portegies Zwart} S.,  2012, \mn@doi [Astron.
  Astrophys.] {10.1051/0004-6361/201218966}, 546, A70

\bibitem[\protect\citeauthoryear{Vandenbroucke \& Wood}{Vandenbroucke \&
  Wood}{2019}]{Vandenbroucke2019}
Vandenbroucke B.,  Wood K.,  2019, \mn@doi [MNRAS] {10.1093/mnras/stz1841},
  488, 1977

\bibitem[\protect\citeauthoryear{Walch \& Naab}{Walch \&
  Naab}{2015}]{Walch2015}
Walch S.,  Naab T.,  2015, \mn@doi [MNRAS] {10.1093/mnras/stv1155}, 451, 2757

\bibitem[\protect\citeauthoryear{{de Avillez}}{{de
  Avillez}}{2000}]{deAvillez2000}
{de Avillez} M.~A.,  2000, \mn@doi [\mnras] {10.1046/j.1365-8711.2000.03464.x},
  \href {https://ui.adsabs.harvard.edu/abs/2000MNRAS.315..479D} {315, 479}

\makeatother
\end{thebibliography}




\appendix


\bsp    
\label{lastpage}
\end{document}